%% file: Proper_Conflict-free_Coloring_arXiv.tex
\documentclass[a4paper]{article}
\pdfoutput=1
\usepackage{graphicx,wrapfig}
\usepackage{lipsum}
\usepackage{amsmath,amssymb,mathrsfs}
\usepackage{hyperref}
\usepackage{url}
\usepackage{mathtools}
\usepackage{changepage}
\usepackage{multirow}
\usepackage{wrapfig}
\usepackage{subfig}
\usepackage{xcolor}
\usepackage{epsfig,enumerate,amsmath,amsfonts,amssymb,amsthm,mathrsfs,ifpdf}
\usepackage{indentfirst,relsize}
\usepackage{setspace,graphicx}
\usepackage{latexsym}
\usepackage[all]{xy}
\usepackage[usenames,dvipsnames]{pstricks}
\usepackage{pst-grad} 
\usepackage{pst-plot} 
\usepackage[margin = 2.50cm]{geometry}
\usepackage[T1]{fontenc}
\usepackage[utf8]{inputenc} 

\usepackage[linesnumbered,ruled,vlined]{algorithm2e}

\usepackage{color,soul}

\usepackage[colorinlistoftodos]{todonotes}

\usepackage{authblk} 

\usepackage{orcidlink}

\usepackage{enumitem}

\usepackage{tikz}

\newcommand{\pcf}{\chi_{\mathrm{pcf}}}

\newcommand{\remove}[1]{}

\newtheorem{theorem}{Theorem}
\newtheorem{lemma}[theorem]{Lemma}
\newtheorem{proposition}[theorem]{Proposition}
\newtheorem{corollary}[theorem]{Corollary}

\newtheorem{conjecture}{Conjecture}

\newcounter{Case}[theorem]

\newcounter{Ca}[theorem]

\newcounter{NCa}[theorem]

\newtheorem{definition}[theorem]{Definition}

\usepackage{xcolor}
\allowdisplaybreaks

\title{Proper Conflict-Free Coloring on Structured Graph Classes: Bounds, Exact Algorithms, and Hardness}

\author[]{Kamal Santra \orcidlink{0009-0006-5997-1452} \footnote{kamal.7.2013@gmail.com, kamal.santra@iitg.ac.in}}
\affil[]{Department of Mathematics\\
	
	Indian Institute of Technology Guwahati\\
	
	Guwahati, 781039, Assam, India}
\date{}

\begin{document}

\maketitle
\begin{abstract}
	A proper conflict-free coloring is a proper vertex coloring in which every
	non-isolated vertex has a neighbor whose color appears exactly once in its
	open neighborhood. We study this coloring on chordal and ordered bipartite
	graph classes. For well-partitioned chordal graphs, we prove the tight bound
	$\omega(G)\leq\pcf(G)\leq\omega(G)+1$ and give an $O(n^2)$ construction
	from a partition tree. Every assignment of lists of at least $\omega(G)+1$
	colors also admits a PCF coloring that respects the lists.
	Deciding whether $\omega(G)$ colors suffice is
	NP-complete, even when the partition tree is a star. This also proves
	NP-completeness on chordal graphs when the number of colors is part of the
	input. For interval graphs, we prove that lists of size $\omega(G)+1$
	always suffice, improving the previous ordinary-coloring bound of
	$\omega(G)+2$. We use matchings to characterize when $\omega(G)$ colors
	suffice and give an endpoint algorithm taking
	$2^{O(\omega(G)^2)}n^{O(1)}$ time. We also characterize the exact value
	when the clique number is at most three and give a linear-time algorithm
	from sorted interval endpoints. For biconvex graphs, we prove a tight
	five-color bound and give a direct exact algorithm taking $O(n^{33})$ time.
	For bipartite permutation graphs, a smaller dynamic program gives an exact
	$O(n^{17})$ algorithm. Finally, if a bipartite graph is convex over an
	ordered shore $X$, we prove the sharp bound
	$\pcf(G)\leq\lceil\log_2(|X|+1)\rceil+3$ and construct a coloring in
	linear time from the order.
\end{abstract}


\section{Introduction}
\label{sec:introduction}

A proper coloring gives different colors to adjacent vertices. In a proper
conflict-free coloring, each non-isolated vertex must also have a neighbor
whose color appears only once in its open neighborhood. We call this
neighbor a witness. Proper coloring of the square of a graph satisfies this
condition, since all neighbors of a vertex then have different colors. A
proper conflict-free coloring asks for much less: one uniquely colored
neighbor is enough. If colors are viewed as communication channels, the two
requirements have simple roles. Properness separates adjacent transmitters,
and the witness condition lets every vertex identify at least one neighbor
by its color.

Even a path on three vertices shows why the extra condition matters. A
proper two-coloring gives both leaves the same color, so the middle vertex
has no witness. A third color is needed. On larger graphs, the gap between
ordinary and proper conflict-free coloring can be much larger. We study
both the number of additional colors needed and the difficulty of finding
an optimal coloring.

Fabrici et al. introduced proper
conflict-free coloring in its present form
\cite{FabriciLuzarRindosovaSotak2023}. Caro et al. studied its relation to ordinary coloring and maximum degree
\cite{CaroPetrusevskiSkrekovski2023}. They conjectured that every connected
graph with maximum degree at least three has a proper conflict-free coloring
with at most $\Delta+1$ colors. Cho et al. proved the
conjecture for chordal graphs \cite{ChoChoiKwonPark2025}. Liu obtained
general list-coloring bounds using structural graph parameters
\cite{Liu2024}.

The computational problem is difficult even on bipartite graphs, where
ordinary coloring needs at most two colors. Ahn et al. proved
NP-completeness for every fixed number of colors $k\geq3$
\cite{AhnImOum2025}. Pradhan and Sharma proved hardness on
perfect-elimination bipartite graphs and gave algorithms for block, proper
interval, chain, and pseudo-split graphs \cite{PradhanSharma2026}.
Sharma et al. proved the split-graph bound
$\chi(G)\leq\pcf(G)\leq\chi(G)+1$ and gave exact equality criteria
\cite[Theorems~1--3, pp.~503--504]{SharmaPaulPandey2026}.
Their AT-free bound $\pcf(G)\leq\chi(G)+3$ implies an additive-three
bound for permutation graphs and a tight five-color bound for bipartite
permutation graphs
\cite[Theorem~9, Corollaries~4--5, and Lemma~1, pp.~507--508]{SharmaPaulPandey2026}.
For interval graphs, they proved $\pcf(G)\leq\chi(G)+2$
\cite[Theorem~11, p.~509]{SharmaPaulPandey2026}.
For proper interval graphs, they proved the sharp $\omega(G)+1$ bound,
an equality criterion, and linear-time exact solvability
\cite[Theorems~12--13 and Corollary~3, p.~510]{SharmaPaulPandey2026}.
Jim\'enez et al. proved that six colors suffice for biconvex graphs, whereas
no constant suffices for convex bipartite graphs \cite{JimenezEtAl2026}.

\paragraph{The graph classes considered.}
Chordal graphs are perfect, so their ordinary chromatic number is their
clique number. This makes them a natural setting for comparing the two
coloring parameters. We first consider well-partitioned chordal graphs.
Such a graph has a partition into clique bags arranged in a tree. Edges
between adjacent bags form a complete bipartite graph between two boundary
sets. The class contains split and block graphs
\cite{AhnJaffkeKwonLima2022}. This representation lets us color one bag at
a time while keeping a witness color unique. It also allows large leaf
cliques. Our hardness proof shows that these leaf cliques are enough to
make exact coloring difficult, even when the partition tree is a star.

We next consider interval graphs. Their vertices are intervals on a line,
and the intervals containing a fixed point form a clique. We use this order
to construct a coloring from left to right. Interval graphs also lie
between proper interval graphs, for which exact algorithms are known, and
chordal graphs, for which we prove hardness when the number of colors is
part of the input. Neither result settles exact coloring on general
interval graphs.

Our last setting is a hierarchy of bipartite graph classes. Every bipartite
permutation graph is biconvex, and every biconvex graph is convex bipartite.
In a biconvex graph, both parts can be ordered so that every neighborhood
is consecutive. A connected graph in this class also has a multichain
distance layering, which we use for coloring. A convex bipartite graph
requires a consecutive order on only one shore. Ordinary coloring uses at
most two colors throughout this hierarchy. Proper conflict-free coloring
behaves differently: five colors suffice in the biconvex case, while the
number needed in the convex case is unbounded. We give a quantitative bound
for this larger class.

\paragraph{Our results.}
Write $n=|V(G)|$, $m=|E(G)|$, and $\omega=\omega(G)$. Our results are as
follows.
\begin{enumerate}[leftmargin=*,itemsep=0.45em]
	\item \textbf{Well-partitioned chordal graphs.}
	We prove $\omega\leq\pcf(G)\leq\omega+1$ and give an $O(n^2)$ coloring
	algorithm from a partition tree. The proof also shows that lists of size
	$\omega+1$ suffice. The bound is tight. Deciding whether
	$\omega$ colors suffice is NP-complete, even when the partition tree is a
	star, and all vertices of each leaf bag have the same neighborhood in the
	central bag. The same reduction proves NP-completeness on chordal graphs
	when the number of colors is part of the input. For each fixed $k$, the
	chordal problem is polynomial-time solvable by a standard treewidth dynamic
	program.
	
	\item \textbf{Interval graphs.}
	We prove that arbitrary lists of size $\omega+1$ suffice. Since interval
	graphs are perfect, this improves the earlier $\chi+2$ bound for the
	whole interval class by one color and also allows prescribed lists. The
	ordinary $\omega+1$ bound was already known for proper interval graphs.
	When every vertex
	has the common set of colors $[\omega+1]$, we construct a coloring in
	$O(n\log n)$ time from an interval model. To study whether $\omega$ colors
	suffice, we give a chain-cover characterization and a matching criterion
	for prescribed witnesses. An endpoint dynamic program takes
	$2^{O(\omega^2)}n^{O(1)}$ time. We characterize the exact value for
	$\omega\leq3$ and give a linear-time algorithm from sorted interval
	endpoints. An eleven-vertex example shows that a simple intersection test
	on maximum cliques does not extend to $\omega=4$.
	
	\item \textbf{Biconvex and bipartite permutation graphs.}
	We prove that five colors suffice for every bipartite graph with a
	multichain ordering. This gives a tight bound for biconvex graphs and
	improves the known bound of six. The five-color bound and a tight example
	for bipartite permutation graphs are due to Sharma et al.
	\cite[Corollary~5 and Lemma~1, pp.~507--508]{SharmaPaulPandey2026}.
	We retain their example and proof to show tightness for the larger
	biconvex class.
	Our direct dynamic programs find an optimal coloring in $O(n^{33})$ time
	on biconvex graphs and $O(n^{17})$ time on bipartite permutation graphs.
	The second algorithm uses a fixed-color dynamic program that applies to
	every permutation graph supplied with a permutation diagram.
	
	\item \textbf{Convex bipartite graphs.}
	If $G$ is convex over an ordered shore $X$, we prove
	$\pcf(G)\leq\lceil\log_2(|X|+1)\rceil+3$. Given the order, a coloring
	can be found in $O(n+m)$ time. We construct an infinite family attaining
	equality. Unboundedness on this class was already known
	\cite{JimenezEtAl2026}; our result gives a sharp bound in terms of $|X|$.
\end{enumerate}

\paragraph{Relation to general algorithms.}
General results already imply polynomial-time solvability on biconvex
graphs and bipartite permutation graphs. For a fixed number of colors,
proper conflict-free coloring is a locally checkable condition that needs
only color counts truncated at two. The count of a vertex's own color must
be zero, and a non-isolated vertex must see a count of one. The algorithm
of Gonzalez and Mann \cite{GonzalezMann2024}, together with the
neighborhood decompositions of Belmonte and Vatshelle
\cite{BelmonteVatshelle2013} and the known constant biconvex bound, therefore
gives an exact polynomial-time algorithm. Our dynamic programs give direct
constructions and explicit running-time bounds. The algorithms parameterized
by clique number likewise use standard bounded-treewidth ideas once a bound
on the number of colors is available.

The contributions developed here are the list bounds for well-partitioned
chordal and interval graphs, hardness on well-partitioned chordal graphs,
the interval structural criteria and endpoint algorithm, the extension of
the five-color bound to biconvex graphs, and the sharp convex bound.
We distinguish these results from the known split and proper interval
special cases and from the published bipartite permutation bound and
example. The direct algorithms provide explicit recurrences and running
times in settings where earlier general methods already imply tractability.

\paragraph{Organization.}
Section~\ref{sec:preliminaries} gives notation and the graph
representations. Section~\ref{sec:wpc} treats well-partitioned chordal
graphs and derives the chordal hardness result.
Section~\ref{sec:interval} studies interval graphs.
Section~\ref{sec:ordered-bipartite-pcf} treats the ordered bipartite
classes. Section~\ref{sec:conclusion} gives further questions.

\section{Preliminaries}
\label{sec:preliminaries}

All graphs are finite, simple, undirected, and nonempty unless stated otherwise.
We write $n=|V(G)|$ and $m=|E(G)|$. We use $N_G(v)$ and $N_G[v]=N_G(v)\cup\{v\}$
for the open and closed neighborhoods of $v$, $\Delta(G)$ for the maximum
degree, and $\omega(G)$ for the clique number. We omit the subscript when
the graph is clear. A stable set contains no adjacent pair. For a positive
integer $k$, let $[k]=\{1,\ldots,k\}$. The empty graph has PCF chromatic
number zero; a nonempty edgeless graph has PCF chromatic number one. Empty
graphs may be handled separately in all algorithms.

\begin{definition}
	A coloring $c:V(G)\to[k]$ is a \emph{proper conflict-free coloring}, or a
	\emph{PCF coloring}, if it is proper and, for every non-isolated vertex $v$,
	there is a color $i$ with $|N(v)\cap c^{-1}(i)|=1$. The unique neighbor
	with that color is a \emph{witness} of $v$. The least number of colors in a
	PCF coloring is denoted by $\pcf(G)$.
\end{definition}

For a list assignment $L$, a PCF $L$-coloring must also satisfy
$c(v)\in L(v)$ for every vertex. A graph is PCF $k$-choosable if such a
coloring exists for every assignment of lists of size at least $k$.
The problem \textsc{PCF Colorability} takes $(G,k)$ and asks whether
$\pcf(G)\leq k$. In \textsc{PCF-$\omega$ Colorability}, the target is the
clique number of the input graph. We distinguish the case in which $k$ is fixed from the case in which it
is part of the input.

The PCF condition depends only on whether a color appears zero times, once,
or at least twice in a neighborhood. We therefore replace every relevant
count $a$ by $\min\{2,a\}$. All algorithms check the condition only at
non-isolated vertices. Different components may reuse the same colors.
An algorithm is fixed-parameter tractable in a parameter $p$ if its running
time is $f(p)n^{O(1)}$. A bound $n^{f(p)}$ gives an XP algorithm in $p$.

\paragraph{Chordal and well-partitioned chordal graphs.}
A graph is \emph{chordal} if it has no induced cycle of length at least
four. Chordal graphs are perfect, have a perfect elimination ordering, and
have treewidth $\omega(G)-1$. A connected graph is
\emph{well-partitioned chordal} if its vertices can be partitioned into
clique bags that form a tree $\mathcal T$. Nonadjacent bags are
anticomplete. Between adjacent bags $X$ and $Y$, the edges are exactly
$X_Y\times Y_X$ for subsets $X_Y\subseteq X$ and $Y_X\subseteq Y$.
These subsets are the \emph{boundaries} of the tree edge. A partition tree
can be found in polynomial time \cite{AhnJaffkeKwonLima2022}.

\paragraph{Interval graphs.}
An \emph{interval graph} is the intersection graph of closed intervals
$I_v=[\ell_v,r_v]$ on the real line. Endpoints can be made distinct without
changing the graph: at an equal coordinate, place all left endpoints before
all right endpoints, then order ties within each type arbitrarily. The
maximal cliques have a \emph{clique-path order} in which the cliques
containing each vertex are consecutive. The \emph{interval order} places
$u$ before $v$ when $r_u<\ell_v$. Stable sets are exactly its chains.

\paragraph{Ordered bipartite classes.}
A bipartite graph $G=(X,Y;E)$ is \emph{convex over $X$} if $X$ has an
order in which $N(y)$ is consecutive for every $y\in Y$. It is
\emph{biconvex} if both shores have such an order. A \emph{bipartite
	permutation graph} has orders $<_X$ and $<_Y$ such that, whenever
$x<_Xx'$, $y<_Yy'$, and $xy',x'y\in E(G)$, both $xy$ and $x'y'$ are
edges. These are \emph{strong orders}. Recognition and the construction of
strong orders take linear time \cite{SpinradBrandstadtStewart1987}.

A distance layering $L_0,L_1,\ldots,L_z$ is a \emph{multichain ordering}
if the edges between each pair of consecutive layers form a chain graph.
A chain graph is bipartite, and its neighborhoods on each shore are nested
by inclusion. Every connected biconvex graph has a multichain ordering
\cite{DiazDinerSernaSerra2021}. We apply this property only to bipartite
graphs, so all edges join consecutive layers.

\input{pcf_well_partitioned}
\input{pcf_interval}

\input{pcf_biconvex}

\section{Conclusion and future research}
\label{sec:conclusion}

We proved tight coloring bounds from three graph representations. A
partition tree gives a coloring with at most one color beyond the clique
number on well-partitioned chordal graphs. The argument also works for
lists of size $\omega+1$. An interval model gives the same list bound on
interval graphs, improving the earlier ordinary $\omega+2$ bound. A
multichain ordering extends the known bipartite permutation bound to a
five-coloring on biconvex graphs, while one-sided convexity gives a sharp
logarithmic bound. The complexity of exact coloring differs across these
classes. Deciding whether the clique bound can be attained is NP-complete
on well-partitioned chordal graphs. Biconvex and bipartite permutation graphs
admit direct polynomial-time algorithms.

The following questions arise from these results.
\begin{enumerate}[leftmargin=*,itemsep=0.4em]
	\item \textbf{Exact coloring of interval graphs.}
	Is deciding whether $\pcf(G)=\omega(G)$ polynomial-time solvable or
	NP-complete? Once witnesses are prescribed, the matching criterion gives a
	polynomial test. Choosing the witnesses remains the main difficulty. A
	structural characterization for $\omega=4$ would be useful: the
	eleven-vertex example shows that singleton intersections of maximum cliques
	are not enough. For every fixed clique number, the endpoint algorithm is
	already polynomial. Any NP-hardness reduction must therefore use unbounded
	clique number.
	
	\item \textbf{A common part of all maximum cliques.}
	Conjecture~\ref{conj:int-maximum-core} asks whether two vertices common to
	all maximum cliques guarantee a PCF $\omega$-coloring of a connected
	interval graph. Another question is to describe the minimal graphs needing
	$\omega+1$ colors when $\omega\geq4$. Such a description must account for
	the fact that an induced subgraph may lack a witness even when the whole
	graph has one.
	
	\item \textbf{Faster direct algorithms.}
	Can the exponents $33$ and $17$ be reduced? The current algorithms store at
	most two representatives of each color in each direction. It may be
	possible to combine states that have the same effect on later vertices.
	Structural characterizations of graphs admitting PCF colorings with three
	or four colors could also lead to simpler algorithms.
	
	\item \textbf{Exact coloring of convex bipartite graphs.}
	Determine the complexity when the number of colors is part of the input.
	Combining the logarithmic bound with the general algorithm of Gonzalez and
	Mann \cite{GonzalezMann2024} gives an exact
	$n^{O(\log n)}$-time algorithm. It remains open whether this can be improved
	to polynomial time, or whether the problem is NP-hard on convex bipartite
	graphs.
	
	\item \textbf{List coloring on other chordal classes.}
	Both interval graphs and well-partitioned chordal graphs admit PCF colorings
	from lists of size $\omega+1$. Which further chordal subclasses have this
	property? The two proofs use different features: an interval order in one
	case and complete bipartite boundaries in the other. A condition that
	includes both constructions could give a common explanation for the bounds.
\end{enumerate}

\bibliographystyle{plain}
\bibliography{pcf_ref}
         
\end{document}

%% file: pcf_well_partitioned.tex
\section{Well-partitioned chordal graphs}
\label{sec:wpc}

Sharma et al. proved that split graphs satisfy
$\chi_{\mathrm{pcf}}(G)\leq\chi(G)+1$
\cite[Theorem~1, p.~503]{SharmaPaulPandey2026}. We extend this bound to
well-partitioned chordal graphs: $\omega(G)+1$ colors always suffice.
This leaves only two possible values for the PCF chromatic number.
We then prove that deciding whether $\omega(G)$ colors suffice is
NP-complete, even when a simple partition tree is given. The same
reduction proves hardness on chordal graphs when the number of colors
is part of the input.

\subsection{Partition trees and boundaries}

A connected graph is \emph{well-partitioned chordal} if its vertices can
be partitioned into cliques, called bags, that are arranged as the nodes
of a tree $\mathcal T$. Bags that are not adjacent in $\mathcal T$ have
no edges between them. Between adjacent bags $X$ and $Y$, all edges run
between a subset of $X$ and a subset of $Y$, and every possible edge
between these subsets is present. The tree $\mathcal T$ is a
\emph{partition tree}. Ahn et al.introduced
this class and gave a polynomial-time recognition algorithm that produces
a partition tree~\cite{AhnJaffkeKwonLima2022}. For a disconnected graph,
we use a partition tree for each component.

For adjacent bags $X,Y$, write
$X_Y=\{x\in X:N_G(x)\cap Y\neq\varnothing\}$ and
$Y_X=\{y\in Y:N_G(y)\cap X\neq\varnothing\}$. These are the
\emph{boundaries} of the edge $XY$. Thus, the edges between the two bags
are exactly $X_Y\times Y_X$. Both bags are cliques, so $X_Y\cup Y_X$ is
also a clique. In particular, $|X_Y|+|Y_X|\leq\omega(G)$.

A clique meets at most two bags. Otherwise, three bags containing its
vertices would be pairwise adjacent, which is impossible in a tree.
If a clique meets two bags, these bags are adjacent, and its vertices lie
in their boundaries. Conversely, each bag and each union of two opposite
boundaries is a clique. Therefore,
\begin{equation}
 \omega(G)=\max\left\{
       \max_{X\in\mathcal P}|X|,
       \max_{XY\in E(\mathcal T)}\bigl(|X_Y|+|Y_X|\bigr)
       \right\}.
 \label{eq:wpc-clique-number}
\end{equation}
Here $\mathcal P$ is the set of bags. If the tree has only one bag, we
omit the second maximum. This formula lets us compute the clique number
from the supplied partition tree and its boundaries.

\subsection{A coloring with one additional color}

Properness gives the lower bound $\omega(G)$. To obtain the upper bound,
we color the bags one at a time, starting at a leaf of the partition tree.
In each bag with at least two vertices, we choose two vertices to serve
as witnesses. We call them anchors. The first anchor will witness every
other vertex of its bag, and the second will witness the first. When we
color a child bag, we keep the relevant witness colors out of its boundary.

\begin{theorem}
\label{thm:wpc-upper-bound}
Every well-partitioned chordal graph $G$ satisfies
$\omega(G)\leq\chi_{\mathrm{pcf}}(G)\leq\omega(G)+1$.
Given a partition tree for each component, a PCF coloring with at most
$\omega(G)+1$ colors can be found in $O(n^2)$ time.
\end{theorem}

\begin{proof}
We may consider one connected component at a time and reuse the colors in
other components. An isolated vertex can receive any color. Consider,
therefore, a connected graph with at least two vertices. Put
$\omega=\omega(G)$, $K=\omega+1$, and $\Gamma=[K]$.
We will give distinct colors to the vertices within each bag.

If the partition tree has one bag, the graph is a clique and any coloring
with distinct colors is PCF. Otherwise, root the tree at a leaf bag and
process parents before their children. Connectedness ensures that both
boundaries of each tree edge are nonempty.

For a bag $X$ with at least two vertices, let $a_X,b_X$ denote its two
anchors. Our goal is to make $a_X$ a witness for every vertex of
$X\setminus\{a_X\}$ and to make $b_X$ a witness for $a_X$.
If a nonroot bag is a singleton $X=\{x\}$, we choose its witness $w_X$
in the parent boundary instead. We choose these vertices as the bags are
colored.

Color the root bag with distinct colors. If it has at least two vertices,
choose any two as its anchors. If the root is a singleton, it needs no
anchor at this stage. It has only one child, because we rooted the tree
at a leaf, and any vertex in that child's boundary will witness the root.

Now let $X$ be an uncolored bag whose parent $P$ has already been colored.
Put $R=X_P$ and $S=P_X$. All edges between these bags join $R$ to $S$.
We must therefore avoid the colors of $S$ when coloring $R$. We also
protect at most one further color. If $P$ has two anchors, define
\begin{equation}
 \eta(P,X)=
 \begin{cases}
 c(a_P),& a_P\notin S,\\
 c(b_P),& a_P\in S\text{ and }b_P\notin S,\\
 \bot,& a_P,b_P\in S.
 \end{cases}
 \label{eq:wpc-protected-color}
\end{equation}
The symbol $\bot$ means that there is no additional color to protect.
If $P$ is a nonroot singleton, put $\eta(P,X)=c(w_P)$.
If $P$ is the singleton root, put $\eta(P,X)=\bot$.
Let $F(P,X)$ consist of the colors on $S$, together with $\eta(P,X)$
when it is defined.

The additional color, when present, is different from every color on $S$.
For a parent with two anchors, the protected anchor lies outside $S$,
and the colors in $P$ are distinct. For a singleton parent, $S=P$ and
$w_P$ is adjacent to its only vertex, so properness gives the same
conclusion. In all cases, $|F(P,X)|\leq |S|+1$.
Since $R\cup S$ is a clique, we have $|R|+|S|\leq\omega$.
Thus at least $K-|S|-1=\omega-|S|\geq |R|$ colors remain for $R$.
Give its vertices distinct colors outside $F(P,X)$.

If $X$ is a singleton, choose any $w_X\in S$ and proceed to the next bag.
Suppose instead that $|X|\geq2$. Choose $a_X\in R$. If $|R|\geq2$,
choose $b_X\in R\setminus\{a_X\}$. Both anchor colors then avoid all
colors on $S$. If $R=\{a_X\}$, choose $b_X\in X\setminus R$ and give
it a color outside $c(S)\cup c(R)$. This is possible because
$K-|S|-1=\omega-|S|\geq1$. In either case, both anchor colors are
absent from the parent boundary. Give distinct unused colors to the
remaining vertices of $X$. This is possible because $|X|\leq\omega<K$.

We first check properness. Vertices in one bag have different colors.
For adjacent bags, the colors on the child boundary avoid the colors on
the parent boundary. These are all the edges of the graph, so the coloring
is proper.

We next check the witnesses. Let $X$ have at least two vertices and take
$x\in X\setminus\{a_X\}$. Inside $X$, the color of $a_X$ occurs only
once. If $X$ is not the root and $x$ has a neighbor in its parent, that
neighbor lies in $P_X$. The color of $a_X$ was chosen outside $c(P_X)$,
so the parent does not duplicate it. Now consider a child $Y$ adjacent
to $x$. If $a_X\in X_Y$, properness keeps its color out of $Y_X$.
If $a_X\notin X_Y$, the first rule in
\eqref{eq:wpc-protected-color} protects its color. Hence no child
duplicates the color either, and $a_X$ witnesses $x$.

For the vertex $a_X$, use $b_X$ as the witness. Its color occurs only
once in $X$ and, when a parent exists, was chosen outside the parent
boundary. A child $Y$ has no effect if $a_X\notin X_Y$.
If both anchors belong to $X_Y$, properness keeps the color of $b_X$
out of $Y_X$. If only $a_X$ belongs to $X_Y$, the second rule in
\eqref{eq:wpc-protected-color} protects the color of $b_X$.
Thus $b_X$ witnesses $a_X$.

Finally, let $X=\{x\}$ be a nonroot singleton. Its chosen witness
$w_X$ belongs to the parent boundary. The colors in the parent bag are
distinct, so $c(w_X)$ occurs exactly once among the neighbors of $x$
in that bag. Every child boundary avoids $c(w_X)$ by the singleton rule.
Thus $w_X$ remains a witness in the completed graph. If the root itself
is a singleton, its neighbors are exactly the vertices of its only
child's boundary. Those vertices have distinct colors, so the root also
has a witness. The coloring is therefore PCF.

The clique number is computed from \eqref{eq:wpc-clique-number}.
Each bag is processed once. Scanning the palette when assigning a color
to a vertex takes at most $O(n)$ time. The total size of all boundary
lists is at most $O(n^2)$, so forming the forbidden sets also fits within
$O(n^2)$ time. The algorithm stores the coloring and the chosen witnesses.
The lower bound follows from properness.
\end{proof}

Algorithm~\ref{alg:wpc-pcf} gives the construction for one connected
component. Its input includes the boundary sets.

\begin{algorithm}[t]
\DontPrintSemicolon
\KwIn{A connected well-partitioned chordal graph $G$ and a partition tree
$\mathcal T$ with all boundaries.}
\KwOut{A PCF coloring $c$ using at most $\omega(G)+1$ colors.}
Compute $\omega$ from Equation~\eqref{eq:wpc-clique-number} and set
$\Gamma\leftarrow[\omega+1]$\;
\If{$\mathcal T$ has one bag $X$}{
  color $X$ injectively and \KwRet{$c$}\;
}
Choose a leaf bag $X_0$ as the root and color $X_0$ injectively\;
\If{$|X_0|\geq2$}{choose distinct anchors $a_{X_0},b_{X_0}\in X_0$\;}
\ForEach{non-root bag $X$ in root-to-leaf order}{
  let $P$ be the parent of $X$, and put $R\leftarrow X_P$ and
  $S\leftarrow P_X$\;
  \eIf{$|P|\geq2$}{
    \uIf{$a_P\notin S$}{$\eta\leftarrow c(a_P)$\;}
    \uElseIf{$b_P\notin S$}{$\eta\leftarrow c(b_P)$\;}
    \Else{$\eta\leftarrow\bot$\;}
  }{
    \eIf{$P\neq X_0$}{$\eta\leftarrow c(w_P)$\;}
                         {$\eta\leftarrow\bot$\;}
  }
  $F\leftarrow c(S)$; if $\eta\neq\bot$, set
  $F\leftarrow F\cup\{\eta\}$\;
  color $R$ injectively with colors from $\Gamma\setminus F$\;
  \eIf{$|X|=1$}{
    let $x$ be the vertex of $X$ and choose $w_X\in S$\;
  }{
    choose $a_X\in R$\;
    \eIf{$|R|\geq2$}{
      choose $b_X\in R\setminus\{a_X\}$\;
    }{
      choose $b_X\in X\setminus R$ and color it from
      $\Gamma\setminus(c(S)\cup c(R))$\;
    }
    color all still uncolored vertices of $X$ injectively with colors unused
    in $X$\;
  }
}
\KwRet{$c$}\;
\caption{PCF coloring from a partition tree.}
\label{alg:wpc-pcf}
\end{algorithm}

\begin{corollary}
\label{cor:wpc-list}
Every well-partitioned chordal graph $G$ is PCF
$(\omega(G)+1)$-choosable.
\end{corollary}

\begin{proof}
Put $\omega=\omega(G)$ and suppose that each vertex $v$ has a list
$L(v)$ of at least $\omega+1$ colors. We use the same construction as in
Theorem~\ref{thm:wpc-upper-bound}, choosing every color from the list of
the vertex being colored. The root bag can be colored with distinct list
colors, because it has at most $\omega$ vertices.

For a child bag $X$ with parent $P$, put $R=X_P$ and $S=P_X$ as before.
The forbidden set $F(P,X)$ has at most $|S|+1$ colors. Thus each vertex
$x\in R$ has at least $\omega-|S|\geq|R|$ colors in
$L(x)\setminus F(P,X)$. Color the vertices of $R$ one at a time, avoiding
the colors already used in $R$. At most $|R|-1$ such colors have been
used at any step, so a choice always remains.

Choose the anchors as in the theorem. If $R$ has at least two vertices,
both anchors are already colored. If $R=\{a_X\}$ and $|X|\geq2$,
choose $b_X\in X\setminus R$. Its list contains a color outside
$c(S)\cup\{c(a_X)\}$, because this set has at most
$|S|+1\leq\omega$ colors. Finally, color each remaining vertex of $X$
from its own list, avoiding all colors already used in $X$. There are
at most $|X|-1\leq\omega-1$ colors to avoid, so this is possible.
Singleton bags use the same witness rule as before.

These choices preserve every property used in the proof of
Theorem~\ref{thm:wpc-upper-bound}: each bag has distinct colors, child
boundaries avoid their forbidden colors, and both anchors avoid the
parent boundary colors. The same witness argument therefore proves that
the resulting list coloring is PCF. Components can be handled separately.
\end{proof}

One additional color is sometimes necessary. The example below is a block
graph and is covered by the characterization of Pradhan and
Sharma~\cite{PradhanSharma2026}. We include the argument because it
shows why a coloring with only the clique number of colors can fail.

\begin{proposition}
\label{prop:wpc-upper-tight}
For every integer $r\geq2$, there is a well-partitioned chordal graph $D_r$
with $\omega(D_r)=r$ and $\chi_{\mathrm{pcf}}(D_r)=r+1$.
\end{proposition}

\begin{proof}
Take two copies of $K_r$ with exactly one common vertex $z$ and no other
edges between them. Use one copy as a bag $X$ and the $r-1$ remaining
vertices of the other copy as a bag $Y$. The two bags form a partition
tree with boundaries $X_Y=\{z\}$ and $Y_X=Y$. Hence the graph is
well-partitioned chordal and has clique number $r$.

In a proper $r$-coloring, each copy of $K_r$ uses every color once.
The color of $z$ is absent from $N(z)$, and each other color occurs twice
there, once in each clique. Thus $z$ has no witness, so $r$ colors do not
suffice. Theorem~\ref{thm:wpc-upper-bound} gives a PCF coloring with
$r+1$ colors.
\end{proof}

The maximum-degree bound in the next corollary is already known for all
chordal graphs~\cite{ChoChoiKwonPark2025}. It also follows directly from
our clique bound on well-partitioned chordal graphs. This is the bound
conjectured by Caro et al.~\cite{CaroPetrusevskiSkrekovski2023}.

\begin{corollary}
\label{cor:wpc-delta-conjecture}
Every connected well-partitioned chordal graph $G$ with at least two vertices
satisfies $\chi_{\mathrm{pcf}}(G)\leq\Delta(G)+1$.
\end{corollary}

\begin{proof}
If $G$ is complete, distinct colors give
$\chi_{\mathrm{pcf}}(G)=|V(G)|=\Delta(G)+1$.
Otherwise, a maximum clique $Q$ has a vertex with a neighbor outside $Q$,
because $G$ is connected and has more than $|Q|$ vertices. That vertex
has at least $|Q|$ neighbors. Hence $\Delta(G)\geq\omega(G)$, and
Theorem~\ref{thm:wpc-upper-bound} gives the result.
\end{proof}

\subsection{The equality problem is NP-complete}

The upper bound leaves one exact question: does the graph have a PCF
coloring with $\omega(G)$ colors? We call this problem
\textsc{PCF-$\omega$-Colorability}. For well-partitioned chordal graphs,
we supply a partition tree as part of the input. The following hardness
result therefore holds even when the representation is already available.

\begin{theorem}
\label{thm:wpc-omega-hardness}
\textnormal{\textsc{PCF-$\omega$-Colorability}} is NP-complete on
well-partitioned chordal graphs. This remains true when the supplied
partition tree is a star, and all vertices of a leaf bag have the same
neighborhood in the central bag.
\end{theorem}

\begin{proof}
The problem is in NP. We can check properness and the conflict-free
condition in polynomial time. We can also compute $\omega(G)$ from the
supplied partition tree using \eqref{eq:wpc-clique-number}.

We reduce from \textsc{Exact Cover by 3-Sets} (\textsc{X3C})
\cite{GareyJohnson1979}. Let $U=\{u_1,\ldots,u_{3q}\}$ and let
$\mathcal S=\{S_1,\ldots,S_m\}$ be three-element subsets of $U$.
We ask whether $q$ of these sets partition $U$. We may assume that
$q\geq1$ and $m\geq q$. If $q=0$, the answer is yes and we may output
$K_2$, whose PCF chromatic number equals its clique number. If $q\geq1$
and $m<q$, the answer is no and we may output $P_3$, whose clique number
is two and whose PCF chromatic number is three.

For each set $S_i$, introduce two vertices $a_i,b_i$, called its
\emph{ports}. Make $C=U\cup\{a_i,b_i:i\in[m]\}$ a clique, and put
$K=|C|=3q+2m$. For each element $u$, define
$H_u=\{u\}\cup\{a_i,b_i:u\in S_i\}$. Add a clique $L_u$ of size
$K-|H_u|$, and make it complete to $H_u$ and anticomplete to
$C\setminus H_u$. Also add a clique $Z$ of size $2m-q$, complete to
$U$ and anticomplete to all ports. The cliques $L_u$ and $Z$ are pairwise
anticomplete. These are all the vertices and edges of the construction.

The bags have positive size. Indeed,
$|H_u|\leq1+2m$ gives $|L_u|\geq3q-1>0$, and
$|Z|=2m-q\geq1$. There are $3q+1$ leaf bags, each of size at most $K$,
so the construction has polynomial size.
Figure~\ref{fig:wpc-x3c-reduction} shows the partition tree. Each line
represents a complete join between the indicated subsets.

\begin{figure}[t]
\centering
\begin{tikzpicture}[
  bag/.style={draw, rounded corners, thick, align=center,
              minimum width=3.25cm, minimum height=1.25cm, fill=white},
  leaf/.style={draw, rounded corners, thick, align=center,
               minimum width=2.5cm, minimum height=0.9cm, fill=white},
  link/.style={thick},
  every node/.style={font=\small}
]
  \node[bag] (C) at (0,0) {$C=U\,\dot\cup\,\{a_i,b_i:i\in[m]\}$\\
                              $|C|=K=3q+2m$};
  \node[leaf] (L1) at (-4.8,2.55) {$L_{u_1}$\\$|L_{u_1}|=K-|H_{u_1}|$};
  \node[leaf] (Lu) at (0,3.05) {$L_u$\\$|L_u|=K-|H_u|$};
  \node[leaf] (Lr) at (4.8,2.55) {$L_{u_{3q}}$\\$|L_{u_{3q}}|=K-|H_{u_{3q}}|$};
  \node[leaf] (Z) at (0,-2.8) {$Z$\\$|Z|=2m-q$};

  \draw[link] (C.north west) -- node[midway,sloped,above,fill=white,
       inner sep=1pt,font=\scriptsize] {$H_{u_1}\times L_{u_1}$}
       (L1.south east);
  \draw[link] (C.north) -- node[midway,right,fill=white,
       inner sep=1pt,font=\scriptsize] {$H_u\times L_u$} (Lu.south);
  \draw[link] (C.north east) -- node[midway,sloped,above,fill=white,
       inner sep=1pt,font=\scriptsize] {$H_{u_{3q}}\times L_{u_{3q}}$}
       (Lr.south west);
  \draw[link] (C.south) -- node[midway,right,fill=white,
       inner sep=1pt,font=\scriptsize] {$U\times Z$} (Z.north);
  \node[align=center, font=\footnotesize] at (0,-4.0)
       {All leaf bags are pairwise anticomplete; ports are anticomplete to $Z$.};
\end{tikzpicture}
\caption{The star partition tree used in the reduction from \textsc{X3C}.
The central bag is the clique $C$.  Each displayed line denotes a complete
join between the indicated boundary in $C$ and the whole leaf bag.  The
omitted bags $L_{u_2},\ldots,L_{u_{3q-1}}$ are constructed in the same way.}
\label{fig:wpc-x3c-reduction}
\end{figure}

The star centered at $C$, with leaves $L_u$ and $Z$, is a partition tree.
The boundaries of $CL_u$ are $H_u$ and $L_u$; those of $CZ$ are $U$
and $Z$. Thus all vertices of a leaf bag have the same neighbors in $C$.
This also gives a direct check of chordality: each vertex of a leaf bag
is simplicial, and deleting all leaf bags leaves the clique $C$.

The central clique has order $K$, as does $H_u\cup L_u$ for every $u$.
The largest clique using $Z$ is $U\cup Z$, of order
$3q+(2m-q)=K-q<K$. No clique meets two leaf bags.
Consequently, $\omega(G)=K$.

Consider any proper $K$-coloring $c$ of this graph. The clique $C$ uses
every color exactly once, so we can name a color by the vertex of $C$
that receives it. Each clique $H_u\cup L_u$ also uses all $K$ colors.
It follows that $c(L_u)=c(C\setminus H_u)$ for every $u$.
The clique $U\cup Z$ forces the vertices of $Z$ to use $2m-q$ distinct
port colors. Exactly $q$ port colors are missing from $Z$.
Let $T$ be the set of ports with these missing colors. Thus $|T|=q$.

For an element vertex $u$, we have
$N_G(u)=(C\setminus\{u\})\cup L_u\cup Z$.
We can now count every color in this neighborhood:
\begin{center}
\begin{tabular}{p{0.64\linewidth}|c}
\hline
vertex of $C$ representing the color & multiplicity in $N_G(u)$\\
\hline
$u$ & $0$\\
an element $v\neq u$ & $2$\\
a port of $S_i$ with $u\notin S_i$ & $2$ or $3$\\
a port $p$ of $S_i$ with $u\in S_i$ and $p\notin T$ & $2$\\
a port $p$ of $S_i$ with $u\in S_i$ and $p\in T$ & $1$\\
\hline
\end{tabular}
\end{center}
The color of another element occurs once in $C$ and once in $L_u$.
A port belonging to a set that does not contain $u$ is outside $H_u$,
so its color occurs in both $C$ and $L_u$, and possibly also in $Z$.
A port belonging to a set that contains $u$ has its color absent from
$L_u$. That color appears only once in $N(u)$ precisely when it is
also absent from $Z$, that is, when the port belongs to $T$.
Therefore $u$ has a witness if and only if
$T\cap\{a_i,b_i:u\in S_i\}\neq\varnothing$.
In other words, at least one missing port color must correspond to a set
containing $u$.

Every other vertex already has a witness in any proper $K$-coloring.
The two ports $a_i,b_i$ witness one another. To see this for $a_i$,
observe that $c(b_i)$ occurs once in $C$. The port $a_i$ has no neighbor
in $Z$. For every leaf bag $L_u$ adjacent to $a_i$, both ports belong
to $H_u$, so the identity $c(L_u)=c(C\setminus H_u)$ excludes
$c(b_i)$ from that bag. The argument for $b_i$ is the same.
For $x\in L_u$, the entire neighborhood
$(L_u\setminus\{x\})\cup H_u$ is a nonempty clique.
For $z\in Z$, the same is true of
$(Z\setminus\{z\})\cup U$.
The colors in each of these neighborhoods are distinct, so these vertices
also have witnesses.

Suppose the \textsc{X3C} instance has an exact cover
$\{S_i:i\in I\}$, where $|I|=q$. Color $C$ with all $K$ colors.
Color each $L_u$ with the colors on $C\setminus H_u$, using each once.
Choose $T=\{a_i:i\in I\}$ and color $Z$ with all port colors except
those on $T$, again using each once. This coloring is proper.
Since the chosen sets cover $U$, the neighborhood calculation gives a
witness for every element vertex. The preceding paragraph gives witnesses
for all other vertices. Thus $G$ has a PCF $K$-coloring.

Conversely, suppose $G$ has a PCF $K$-coloring, and let $T$ be its set of
$q$ missing port colors, identified with their ports in $C$.
The neighborhood calculation shows that
$\mathcal C_T=\{S_i:a_i\in T\text{ or }b_i\in T\}$ covers $U$.
This family contains at most $q$ distinct sets, each with three elements.
Since $|U|=3q$, we obtain
$3q=|U|=|\bigcup_{S\in\mathcal C_T}S|\leq3|\mathcal C_T|\leq3q$.
Equality throughout forces exactly $q$ sets with no overlap. Hence
$\mathcal C_T$ is an exact cover. This proves the reduction.
\end{proof}

The upper bound and the reduction give the following consequence.

\begin{corollary}
\label{cor:wpc-dichotomy}
Every well-partitioned chordal graph $G$ satisfies
$\chi_{\mathrm{pcf}}(G)\in\{\omega(G),\omega(G)+1\}$.
Deciding whether $\chi_{\mathrm{pcf}}(G)=\omega(G)$ is NP-complete,
even when a star partition tree is supplied. Computing the exact value
is therefore NP-hard. Algorithm~\ref{alg:wpc-pcf} uses at most one color
more than an optimal coloring.
\end{corollary}

\subsection{Consequences for chordal graphs and parameterized algorithms}

Every well-partitioned chordal graph is chordal. We therefore obtain the
same hardness when the input is an arbitrary chordal graph and the number
of colors is part of the input.

\begin{corollary}
\label{cor:chordal-pcf-hardness}
Given a chordal graph $G$ and an integer $k$, deciding whether $G$ has a
PCF coloring with at most $k$ colors is NP-complete. The hardness holds
when $k=\omega(G)$, even on well-partitioned chordal graphs.
\end{corollary}

\begin{proof}
A proposed coloring can be checked in polynomial time. For hardness,
apply Theorem~\ref{thm:wpc-omega-hardness} and output $(G,K)$.
The construction has $\omega(G)=K$ and admits a PCF $K$-coloring exactly
when the \textsc{X3C} instance has an exact cover.
\end{proof}

Here the number of colors is $K=3q+2m$, which grows with the original
instance. For a fixed number of colors, a standard dynamic program on a
tree decomposition gives a polynomial-time algorithm.

\begin{proposition}
\label{prop:chordal-fixed-k}
For every fixed positive integer $k$, PCF $k$-colorability can be decided
in polynomial time on chordal graphs.
\end{proposition}

\begin{proof}
If $\omega(G)>k$, the graph has no proper $k$-coloring.
Otherwise, a perfect elimination ordering gives a tree decomposition of
width $\omega(G)-1\leq k-1$. Convert it into a nice tree decomposition
and process it from the leaves towards an empty root.

For each vertex in the current bag, a state stores its color and one
counter for each of the $k$ colors. A counter records the number of
neighbors of that color that have already been forgotten, with every
value at least two replaced by two. The colors of adjacent bag vertices
must differ. Every forgotten vertex must already have passed the PCF
test. A leaf with an empty bag has one empty state.

At an introduce node, choose a color different from those of the new
vertex's neighbors in the bag and initialize its counters to zero.
There cannot be a previously forgotten neighbor of this vertex, by the
connectedness condition for bags in a tree decomposition.
At a forget node for $v$, add the colors of its neighbors that remain
in the bag to its stored counts. These are all its remaining neighbors.
Unless $v$ is isolated, retain the state only if some resulting count
is one. Then add the color of $v$ to the counters of its remaining
neighbors and remove $v$ from the state.
At a join node, the two child states must give the same colors to the
common bag. Add their counters coordinatewise and cap the sums at two.
The forgotten vertex sets in the two subtrees are disjoint, so no neighbor
is counted twice. A nonempty table at the empty root means that a PCF
coloring exists.

A bag has at most $k$ vertices. Each vertex has $k$ possible colors and
$3^k$ possible counter vectors, so there are at most
$(k3^k)^k=2^{O(k^2)}$ states per bag. Even checking all pairs of states
at a join keeps the dependence on $k$ within $2^{O(k^2)}$.
The number of bags and the work on each state are polynomial in the
input size for fixed $k$. This proves the claim.
\end{proof}

For well-partitioned chordal graphs, we need to test only the smaller
of the two possible values.

\begin{proposition}
\label{prop:wpc-fpt-omega}
For an $n$-vertex well-partitioned chordal graph $G$, an optimal PCF
coloring can be found in $2^{O(\omega(G)^2)}n^{O(1)}$ time.
\end{proposition}

\begin{proof}
Put $\omega=\omega(G)$ and run the preceding dynamic program with the
colors $[\omega]$. A state gives each bag vertex a color and a vector
in $\{0,1,2\}^{\omega}$. Its $i$th coordinate counts the forgotten
neighbors of color $i$, capped at two. Introduce, forget, and join nodes
use exactly the transitions above. In particular, a vertex is tested
only when all its neighbors have been accounted for.

Each bag has at most $\omega$ vertices, so there are at most
$(\omega3^\omega)^\omega=2^{O(\omega^2)}$ states per bag.
The join transitions also fit within $2^{O(\omega^2)}n^{O(1)}$ time.
Store one predecessor choice for each retained state. If the table at
the empty root is nonempty, following these choices recovers a PCF
$\omega$-coloring, which is optimal. If the table is empty,
Theorem~\ref{thm:wpc-upper-bound} gives the exact value $\omega+1$,
and Algorithm~\ref{alg:wpc-pcf} constructs an optimal coloring.
\end{proof}

Sharma et al. gave the additive-one bound and exact equality
criteria for split graphs
\cite[Theorems~1--3, pp.~503--504]{SharmaPaulPandey2026}. These criteria
yield a polynomial-time exact algorithm. Pradhan and Sharma also studied
exact coloring on split graphs~\cite{PradhanSharma2026}. A split graph
has a star partition tree whose leaf bags are singletons. Our construction
still has a star partition tree, but its larger leaf cliques force the
sets of colors that they use. These larger bags already make exact PCF
coloring NP-hard.

%% file: pcf_interval.tex
\section{Interval graphs}
\label{sec:interval}

Sharma et al. proved
$\chi_{\mathrm{pcf}}(G)\leq\chi(G)+2$ for interval graphs
\cite[Theorem~11, p.~509]{SharmaPaulPandey2026}. For proper interval
graphs, they already proved the sharp $\omega(G)+1$ bound, an equality
criterion, and a linear-time exact algorithm
\cite[Theorems~12--13 and Corollary~3, p.~510]{SharmaPaulPandey2026}.
We improve the bound for all interval graphs to $\omega(G)+1$ and prove
that arbitrary lists of this size suffice. Since interval graphs satisfy
$\chi(G)=\omega(G)$, this saves one color from the earlier general
interval bound. In particular, every interval graph has PCF chromatic
number either $\omega(G)$ or $\omega(G)+1$. We then study how to decide
which value occurs.

Let $\mathcal I=\{I(v):v\in V(G)\}$ be an interval representation,
where $I(v)=[\ell(v),r(v)]$.  We may move equal endpoints slightly so that
all endpoints are distinct while preserving every intersection.

\subsection{A list-coloring bound}
\label{subsec:interval-list-bound}

We color intervals in order of their left endpoints.  Each vertex other
than the first chooses an earlier neighbor as its witness.  We choose these
neighbors so that each step forbids at most one extra color beyond those
needed for a proper coloring.

\begin{theorem}
\label{thm:interval-list-pcf}
Let $G$ be an interval graph, let $\omega=\omega(G)$, and let $L$ be a list
assignment such that $|L(v)|\geq\omega+1$ for every $v\in V(G)$.  Then $G$
has a proper conflict-free $L$-coloring.  In particular,
$\chi_{\mathrm{pcf}}(G)\leq\omega(G)+1$.
Given an interval representation, the coloring can be found in
$O(n^2+\sum_v|L(v)|)$ time when color names allow constant-time marking.
If every vertex has the common palette $[\omega+1]$, the running time is
$O(n\log n)$.
\end{theorem}

\begin{proof}
We color each connected component separately.  An isolated vertex may
receive any color in its list, so consider a component with at least two
vertices.

Choose an interval $I(s)$ with the smallest right endpoint.  If $u$ is a
neighbor of $s$, then $\ell(u)\leq r(s)$ and $r(u)\geq r(s)$.
Every interval for a neighbor of $s$ therefore contains $r(s)$, so $N(s)$
is a clique.  We may extend the left
endpoint of $I(s)$ to the left of every endpoint in the representation
without creating an edge.  Indeed, an interval disjoint from $I(s)$ cannot
lie to its left, since its right endpoint would then precede $r(s)$.
Order the vertices as $v_1=s,v_2,\ldots,v_n$ by increasing left endpoint
in the resulting model.

For each $i$, let $q_i$ be the largest index $j$ with
$\ell(v_j)<r(v_i)$.  This index records how far $I(v_i)$ extends in the
left-endpoint order.  For $i\geq2$, let $A_i$ be the set of earlier
intervals that are still active when $I(v_i)$ begins.  Thus
$A_i=\{v_h:h<i\text{ and }q_h\geq i\}$.  These are exactly the
earlier neighbors of $v_i$.  Every interval in $A_i\cup\{v_i\}$ contains
$\ell(v_i)$, and hence $|A_i|+1\leq\omega$.  Moreover, $A_i$ is nonempty.
Otherwise all earlier intervals would end before $I(v_i)$ begins.
No earlier vertex could then be adjacent to $v_i$ or any later vertex,
contradicting connectedness.

For $i\geq2$, choose $v_{a_i}\in A_i$ with $q_{a_i}$ as large as
possible.  We call this vertex the \emph{anchor} of $v_i$; it will be the
witness of $v_i$.  Color the vertices in the order $v_1,\ldots,v_n$.  Give $v_1$ any color in its list.  When
$v_i$ is reached, forbid the colors already appearing on $A_i$.  In
addition, suppose an active vertex $v_h\in A_i$ has an anchor that is no
longer active, so $q_{a_h}<i$.  We also forbid the color of that anchor.
Thus the forbidden set is

\begin{equation}
\label{eq:int-forbidden-set}
 F_i=\{c(v_h):v_h\in A_i\}\cup
 \{c(v_{a_h}):v_h\in A_i,\ a_h\text{ is defined, and }q_{a_h}<i\}.
\end{equation}

At most one active vertex contributes an anchor color to the second set in
\eqref{eq:int-forbidden-set}.  To see this, suppose that $h<h'$ are two
such indices.  Since $v_h\in A_i$, we have $q_h\geq i>h'$, and therefore
$v_h\in A_{h'}$.  The choice of the anchor of $v_{h'}$ gives $q_{a_{h'}}\geq q_h\geq i$.  This contradicts
$q_{a_{h'}}<i$.  The vertices of $A_i$ form a clique, and their colors are
pairwise distinct.  We therefore have $|F_i|\leq |A_i|+1\leq\omega$.
Since the list of $v_i$ contains at least $\omega+1$ colors, a color in
$L(v_i)\setminus F_i$ always exists.  Avoiding the first set of colors makes the coloring proper.

We next show that the anchor of every $v_h$, with $h\geq2$, is a witness.
Since $v_{a_h}$ belongs to the clique $A_h$, properness shows that
$c(v_{a_h})$ occurs exactly once among the earlier neighbors of $v_h$.
Consider any later neighbor $v_i$ of $v_h$.  Then $v_h\in A_i$.  If
$q_{a_h}\geq i$, the anchor is also in $A_i$, and properness prevents
$v_i$ from receiving its color.  If $q_{a_h}<i$, the second set in
\eqref{eq:int-forbidden-set} prevents the same assignment.  Thus $v_{a_h}$ is the only neighbor of $v_h$ with its color.

Finally, $N(v_1)=N(s)$ is a nonempty clique.  All its vertices have distinct
colors in a proper coloring, so any one of them witnesses $v_1$.  This
proves that the constructed coloring is proper and conflict-free.

For arbitrary lists, first keep any $\omega+1$ colors from each list.
The proof applies to these smaller lists as well.  At each step,
scan the active vertices, mark the forbidden colors, and scan the retained
list until an allowed color is found.  With constant-time marking for color
names, this takes $O(n^2)$ time after reading the lists and sorting the
representation.  Including the list input gives
$O(n^2+\sum_v|L(v)|)$ time.  The implementation below improves the bound
when all vertices have the same palette.
\end{proof}

For a common palette, use a max-heap with key $q_h$ to choose each anchor.
A second heap stores $(q_{a_h},h)$ for every active vertex with a defined
anchor, using the first coordinate as a minimum key.  Before consulting
either heap, remove inactive entries from its top.  The second heap then
identifies the one possible anchor color that must still be avoided after
the anchor has expired.
A balanced set stores the colors absent from the active intervals.  Buckets
indexed by $q_h$ identify intervals when they expire.  Algorithm~\ref{alg:interval-greedy}
gives the procedure for one connected component.

\begin{algorithm}[t]
\DontPrintSemicolon
\caption{\textsc{Interval-PCF-Greedy}$(\mathcal I)$}
\label{alg:interval-greedy}
\KwIn{An interval representation of a connected nontrivial interval graph.}
\KwOut{A PCF coloring with colors in $[\omega(G)+1]$.}
Choose an interval $s$ of minimum right endpoint and extend its left endpoint
to the left of all other endpoints\;
Order the intervals as $v_1=s,v_2,\ldots,v_n$ by increasing left endpoint,
compute every $q_i$, and put $K\gets\omega(G)+1$\;
Initialize an empty max-heap $H_q$, an empty min-heap $H_a$, the set
$U\gets[K]$ of currently unused colors, and Boolean flags
$\operatorname{active}[i]\gets\mathrm{false}$\;
Create expiration buckets $D_j=\{h:q_h=j\}$\;
\For{$i\gets1$ \KwTo $n$}{
  \ForEach{$h\in D_{i-1}$}{
    $\operatorname{active}[h]\gets\mathrm{false}$ and insert $c(v_h)$ into
    $U$\;
  }
  Delete inactive entries from the tops of $H_q$ and $H_a$\;
  $\eta\gets\bot$\tcp*{the extra expired-anchor color}
  \If{$i\geq2$}{
    Let $a_i$ be the active index at the top of $H_q$\;
    \If{$H_a$ is nonempty and its top entry $(q_{a_h},h)$ satisfies
    $q_{a_h}<i$}{
      $\eta\gets c(v_{a_h})$\;
    }
  }
  Choose $\gamma\in U\setminus\{\eta\}$ and set $c(v_i)\gets\gamma$\;
  Delete $\gamma$ from $U$, set $\operatorname{active}[i]\gets\mathrm{true}$,
  and insert $(q_i,i)$ into $H_q$\;
  \If{$i\geq2$}{insert $(q_{a_i},i)$ into $H_a$\;}
}
\Return{$c$}\;
\end{algorithm}

The proof of Theorem~\ref{thm:interval-list-pcf} shows that
$U\setminus\{\eta\}$ is never empty.  Each interval enters and leaves the
active structures once, and every heap entry is deleted at most once.
Endpoint sorting and all updates therefore take $O(n\log n)$ time.  If a
sorted endpoint representation is supplied, all $q_i$ and the clique number
can be computed in linear time, but the balanced-set implementation still
gives the stated $O(n\log n)$ bound.

The bound is tight for every clique number at least two. For
$r\geq2$, let $F_r$ consist of two copies of $K_r$ with exactly one common
vertex $z$ and with no edge between their exclusive parts. This is a
proper interval graph with clique number $r$ and maximum degree $2r-2$.
Its value $\chi_{\mathrm{pcf}}(F_r)=r+1$ therefore already follows from
the equality criterion of Sharma, Paul, and Pandey
\cite[Theorem~13, p.~510]{SharmaPaulPandey2026}. We include the short
argument for completeness. In every proper $r$-coloring, both cliques use
every color.  Every color other than $c(z)$ therefore occurs
twice in $N(z)$, while properness excludes $c(z)$ from that neighborhood.
Thus $z$ has no unique-colored neighbor.  Theorem~\ref{thm:interval-list-pcf}
now gives $\chi_{\mathrm{pcf}}(F_r)=r+1$.

Deleting any vertex from $F_r$ reduces its PCF chromatic number to at most
$r$.  Deleting $z$ leaves two copies of $K_{r-1}$.  If a noncentral vertex is deleted, color the intact
$K_r$ with all $r$ colors and color the shortened clique with $r-1$ of
them, retaining the color of $z$.  The color omitted on the shortened side
is unique in $N(z)$, and every other vertex is witnessed inside its clique.

\begin{corollary}
\label{cor:int-dichotomy}
Every interval graph $G$ satisfies
$\chi_{\mathrm{pcf}}(G)\in\{\omega(G),\omega(G)+1\}$, and the upper value
is attained for every $\omega(G)\geq2$.
\end{corollary}

\begin{proof}
Properness gives $\chi_{\mathrm{pcf}}(G)\geq\omega(G)$.  Theorem~\ref{thm:interval-list-pcf}
gives the upper bound, and the graphs $F_r$ attain it.
\end{proof}

\subsection{The exact lower-value problem}
\label{subsec:interval-exact}

We now study when $\omega(G)$ colors suffice.  First, we record how the
answer depends on the connected components.

\begin{lemma}[Components]
\label{lem:int-components}
For a nonempty graph, the PCF chromatic number is the maximum of the values
on its connected components.  An interval graph therefore has value
$\omega(G)+1$ exactly when some component $H$ satisfies
$\omega(H)=\omega(G)$ and $\chi_{\mathrm{pcf}}(H)=\omega(H)+1$.
\end{lemma}

\begin{proof}
A coloring is proper and conflict-free exactly when its restriction to
each component has these properties.  Reusing the palette on different
components proves the first assertion.  For the
second, apply the additive-one theorem to each component and observe that a
component of clique number below $\omega(G)$ uses at most $\omega(G)$
colors.
\end{proof}

For a vertex $v$, let $\mathcal M(v)$ be the family of maximum cliques that
contain $v$.

\begin{lemma}[An obstruction from maximum cliques]
\label{lem:int-maximum-clique-kernel}
Let $v$ be a nonisolated vertex.  If a nonempty subfamily
$\mathcal F\subseteq\mathcal M(v)$ satisfies
$\bigcap_{C\in\mathcal F}C=\{v\}$, then $G$ has no PCF
$\omega(G)$-coloring.  The condition holds if and only if $\mathcal M(v)$ is nonempty and the
intersection of all its members is $\{v\}$.
\end{lemma}

\begin{proof}
Put $k=\omega(G)$ and consider a proper $k$-coloring.  Every clique in
$\mathcal F$ has order $k$ and therefore contains every color exactly once.
Fix a color $\alpha\neq c(v)$.  If $\alpha$ occurred only once in $N(v)$,
say at $x$, then $x$ would be the $\alpha$-colored representative in every
clique of $\mathcal F$.  This would put $x$ in
$\bigcap_{C\in\mathcal F}C=\{v\}$, impossible because
$c(x)=\alpha\neq c(v)$.  Hence every color other than $c(v)$ occurs at
least twice in $N(v)$, while $c(v)$ does not occur there.  Thus $v$ has no
conflict-free witness.
\end{proof}

In particular, two maximum cliques of order at least two that meet in one
vertex force the upper value.  For clique number four, this condition is no
longer necessary, as the following example shows.

\begin{proposition}
\label{prop:int-eleven-counterexample}
There is an interval graph $J$ on eleven vertices such that
$\omega(J)=4$ and $\chi_{\mathrm{pcf}}(J)=5$, although no two maximum
cliques of $J$ meet in exactly one vertex.
\end{proposition}

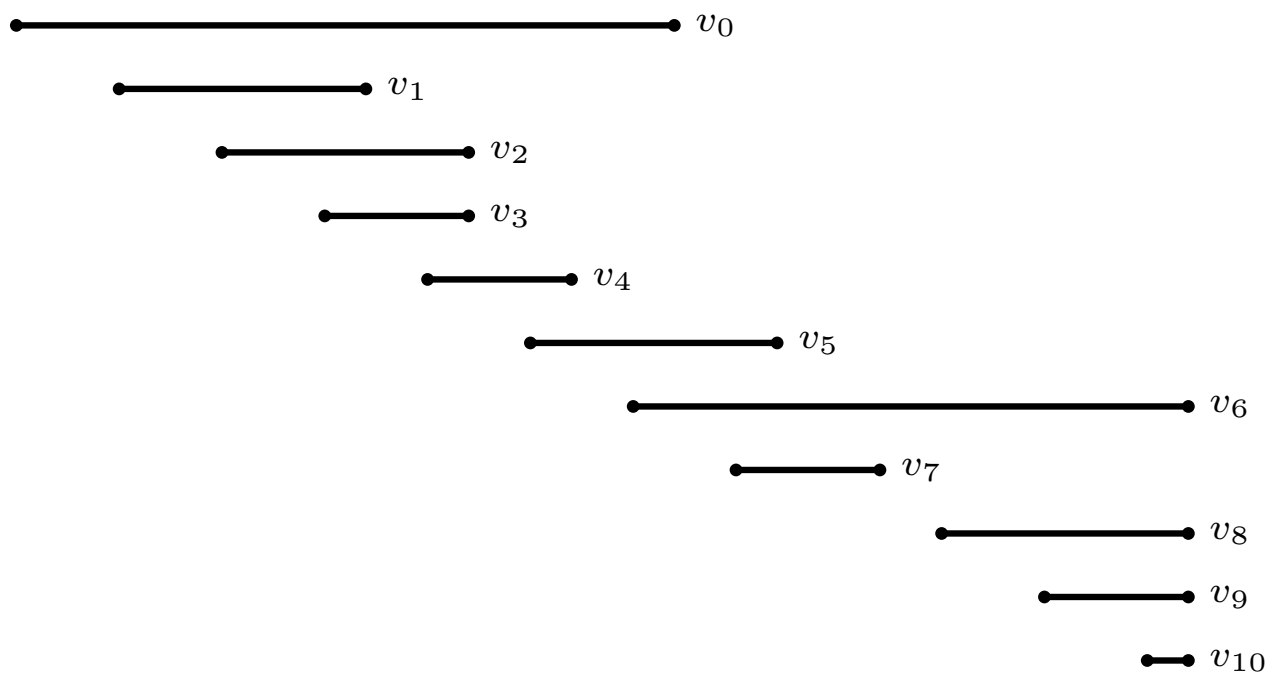
\begin{figure}[t]
\centering
\begin{tikzpicture}[x=0.68cm,y=0.42cm,
  int/.style={very thick,black},
  lab/.style={font=\scriptsize,anchor=west}]
\draw[int] (0,10)--(6.4,10) node[lab] {$v_0$};
\draw[int] (1,9)--(3.4,9) node[lab] {$v_1$};
\draw[int] (2,8)--(4.4,8) node[lab] {$v_2$};
\draw[int] (3,7)--(4.4,7) node[lab] {$v_3$};
\draw[int] (4,6)--(5.4,6) node[lab] {$v_4$};
\draw[int] (5,5)--(7.4,5) node[lab] {$v_5$};
\draw[int] (6,4)--(11.4,4) node[lab] {$v_6$};
\draw[int] (7,3)--(8.4,3) node[lab] {$v_7$};
\draw[int] (9,2)--(11.4,2) node[lab] {$v_8$};
\draw[int] (10,1)--(11.4,1) node[lab] {$v_9$};
\draw[int] (11,0)--(11.4,0) node[lab] {$v_{10}$};
\foreach \x/\y in {0/10,6.4/10,1/9,3.4/9,2/8,4.4/8,3/7,4.4/7,
4/6,5.4/6,5/5,7.4/5,6/4,11.4/4,7/3,8.4/3,9/2,11.4/2,
10/1,11.4/1,11/0,11.4/0}{\fill[black] (\x,\y) circle (1.2pt);}
\end{tikzpicture}
\caption{An interval model of the graph $J$ in
Proposition~\ref{prop:int-eleven-counterexample}.}
\label{fig:int-eleven-counterexample}
\end{figure}

\begin{proof}
Order the vertices $v_0,\ldots,v_{10}$ by left endpoint and use the interval
model shown in Figure~\ref{fig:int-eleven-counterexample}.  Equivalently,
put $(R_0,\ldots,R_{10})=(6,3,4,4,5,7,10,7,10,10,10)$.
For $i<j$, the edge $v_iv_j$ is present
exactly when $j\leq R_i$.  Its only cliques of order four are
$C_1=\{v_0,v_1,v_2,v_3\}$,
$C_2=\{v_0,v_2,v_3,v_4\}$, and
$C_3=\{v_6,v_8,v_9,v_{10}\}$.  Their pairwise intersections have orders
three, zero, and zero.

Suppose that a PCF four-coloring exists.  Name the colors on $C_1$ so that
$c(v_0)=0$, $c(v_1)=1$, $c(v_2)=2$, and $c(v_3)=3$.  The clique $C_2$
forces $c(v_4)=1$.  Since $v_0v_4v_5v_0$ is a triangle, the color of $v_5$
is some $s\in\{2,3\}$; let $u$ denote the other member of $\{2,3\}$.
The vertex $v_6$ is adjacent to $v_0$ and $v_5$, so its color belongs to
$\{1,u\}$.  Now
$N(v_0)=\{v_1,v_2,v_3,v_4,v_5,v_6\}$.  If $c(v_6)=u$, each of the colors
$1,2,3$ occurs at least twice in this neighborhood.  Therefore the
conflict-free requirement at $v_0$ forces $c(v_6)=1$.

The triangle on $v_5,v_6,v_7$ gives $c(v_7)\in\{0,u\}$.  Since $C_3$
is a $K_4$, the colors on $v_8,v_9,v_{10}$ are $0,2,3$ in some order.
Among the neighbors of $v_6$ other than $v_7$, color $0$ occurs on $v_0$
and in $C_3$, color $s$ occurs on $v_5$ and in $C_3$, and color $u$
occurs once, in $C_3$.  The vertex $v_6$ therefore has a witness only if
$c(v_7)=0$.  But then
$N(v_5)=\{v_0,v_4,v_6,v_7\}$ has color multiset
$\{0,1,1,0\}$, a contradiction.

The color vector $(0,1,2,3,1,2,4,0,0,1,2)$ gives a PCF five-coloring.
For completeness, $v_0$ sees color $3$ once, $v_5$ sees color $1$ once,
and $v_6$ sees color $1$ once.  Each of $v_1,v_2,v_3,v_4$ sees the color
of $v_0$ once.  The vertex $v_7$ and each of $v_8,v_9,v_{10}$ see the
color of $v_6$ once.  The displayed vector is proper on every edge, so
$\chi_{\mathrm{pcf}}(J)=5$.
\end{proof}

\subsection{An exact endpoint dynamic program}
\label{subsec:interval-fpt}

The complexity of deciding whether $\omega(G)$ colors suffice remains
open when the clique number is unbounded.  We give an explicit algorithm
whose running time is fixed-parameter tractable in $\omega(G)$.  It follows
the standard dynamic programming approach along the interval endpoints.

Process the endpoints from left to right.  After each endpoint, let $A$ be
the set of intervals that have begun and have not yet ended.  The set $A$
is a clique and has order at most $\omega$.  A state
consists of an injective coloring $c_A:A\to[k]$ and, for every $v\in A$ and
$\alpha\in[k]$, a value $\mu(v,\alpha)\in\{0,1,2\}$.  The value counts already forgotten
neighbors of $v$ with color $\alpha$, with every count of at least two
recorded as $2$.  We retain only partial colorings in which every forgotten
nonisolated vertex has already passed the witness test.

At the left endpoint of $v$, choose a color absent from the active clique,
add $v$ to the state, and set its counters to zero.  This initialization is
correct because an interval that ended before $v$ began cannot be adjacent
to $v$.  At the right endpoint of $v$, every neighbor of $v$ is known.
For each color,
add to its stored counter the number of other active intervals of that
color, again capping at two.  Unless $v$ is isolated, retain the state only
if one of these totals equals one.  Then forget $v$ and increment the
$c_A(v)$-counter of every remaining active interval.  These transitions are
summarized in Algorithm~\ref{alg:interval-fpt}.

\begin{algorithm}[t]
\DontPrintSemicolon
\caption{\textsc{Test-Interval-PCF}$(\mathcal I,k)$}
\label{alg:interval-fpt}
\KwIn{An interval representation $\mathcal I$ and an integer $k$.}
\KwOut{A PCF $k$-coloring, or a report that none exists.}
Sort the $2n$ distinct endpoints of the representation\;
$\mathcal T\gets\{(\varnothing,\varnothing)\}$\;
\ForEach{endpoint event $e$ from left to right}{
  $\mathcal T'\gets\varnothing$\;
  \eIf{$e$ introduces an interval $v$}{
    \ForEach{state $(c_A,\mu)\in\mathcal T$ and each
    $\gamma\in[k]\setminus c_A(A)$}{
      Form a copy of the state, extend $c_A$ by $c_A(v)=\gamma$, and put
      $\mu(v,\alpha)=0$ for every $\alpha\in[k]$\;
      Insert the new state into $\mathcal T'$\;
    }
  }{
    Let $v$ be the interval ending at $e$\;
    \ForEach{state $(c_A,\mu)\in\mathcal T$}{
      For every $\alpha\in[k]$, set
      $t_\alpha\gets\min\{2,\mu(v,\alpha)+
      |\{u\in A\setminus\{v\}:c_A(u)=\alpha\}|\}$\;
      \If{$d_G(v)=0$ or $t_\alpha=1$ for some $\alpha\in[k]$}{
        Work on a copy of the state and put $\gamma\gets c_A(v)$\;
        Delete $v$ from $c_A$ and $\mu$\;
        For every $u\in A\setminus\{v\}$, replace
        $\mu(u,\gamma)$ by $\min\{2,\mu(u,\gamma)+1\}$\;
        Insert the resulting state into $\mathcal T'$\;
      }
    }
  }
  Merge identical states, retain one backpointer for each, and set
  $\mathcal T\gets\mathcal T'$\;
}
\If{$\mathcal T=\varnothing$}{\Return{no PCF $k$-coloring exists}\;}
\Return{a coloring recovered from the backpointers}\;
\end{algorithm}

The counters have the stated meaning after every transition.  When a vertex
is introduced, all its earlier neighbors are active, and the color choices
consider every proper extension.  When a vertex is forgotten, each of its
neighbors has either already been forgotten or is still active.  The test
therefore checks its full neighborhood.  Later intervals cannot affect that
neighborhood.  This proves that every surviving final state gives a PCF
coloring.  Conversely, any PCF $k$-coloring selects a valid color at each
introduction and passes every test at a right endpoint.  It therefore gives
a surviving sequence of states.

For an active set of size $b\leq\omega$, there are at most
$k^b3^{kb}$ states.  With $k=\omega$, this is
$2^{O(\omega^2)}$.  We have therefore proved the following.

\begin{theorem}
\label{thm:int-exact-fpt}
Given an interval representation, one can decide whether
$\chi_{\mathrm{pcf}}(G)=\omega(G)$ and reconstruct a coloring, when one
exists, in $2^{O(\omega(G)^2)}n^{O(1)}$ time.  If the test rejects,
Algorithm~\ref{alg:interval-greedy} constructs an optimal
$(\omega(G)+1)$-coloring.
\end{theorem}

\subsection{Chain covers and matchings}
\label{subsec:interval-chain-cover}

A proper coloring of an interval graph can also be viewed as a partition
into chains. This gives another way to test whether $\omega(G)$ colors
suffice.

Let $P_{\mathcal I}$ be the interval order defined by $u\prec v$ when
$r(u)<\ell(v)$. Two vertices are incomparable exactly when they are
adjacent in $G$. The chains are therefore the stable sets of $G$, and the
antichains are its cliques. The width of $P_{\mathcal I}$ is $\omega(G)$.
By Dilworth's theorem~\cite{Dilworth1950}, a minimum chain cover has
$\omega(G)$ chains. Giving one color to each chain gives a proper
$\omega(G)$-coloring.

Form a bipartite graph $B$ with two copies $V^L,V^R$ of $V(G)$ and an edge
$u^Lv^R$ whenever $u\prec v$. A matching of size $n-\omega(G)$ gives
$\omega(G)$ disjoint directed paths after the two copies of each vertex
are identified. Each vertex has at most one predecessor and one successor.
Every link follows the order, so there is no directed cycle. The paths
form a minimum chain cover. Conversely, linking consecutive vertices of
each chain gives a matching of size $n-\omega(G)$.

For such a matching $M$, let $p_M(w)$ and $s_M(w)$ be the predecessor and
successor of $w$ on its chain, when they exist. Put
$a_M(w)=r(p_M(w))$, with $a_M(w)=-\infty$ when there is no predecessor.
Similarly, put $b_M(w)=\ell(s_M(w))$, with $b_M(w)=+\infty$ when there is
no successor. Call $M$ \emph{locally covering} if every nonisolated vertex
$v$ has a neighbor $w$ such that $a_M(w)<\ell(v)$ and $r(v)<b_M(w)$.
Thus $I(v)$ lies between the predecessor and successor of $w$. It meets
$I(w)$ but no other interval on that chain.

\begin{theorem}[Chain-cover characterization]
\label{thm:int-chain-cover-characterization}
An interval graph $G$ has a PCF $\omega(G)$-coloring if and only if $B$
has a locally covering matching of size $n-\omega(G)$.
\end{theorem}

\begin{proof}
Take a minimum chain cover and a vertex $w$ on one chain with $w\in N(v)$.
Its color occurs only once in $N(v)$ if and only if no other interval of
the chain meets $I(v)$.

We need only inspect the predecessor and successor of $w$. A predecessor
disjoint from $I(v)$ must lie to its left. It cannot lie to the right of
$I(v)$, since it precedes $w$ and $I(w)$ meets $I(v)$. All earlier
members of the chain then lie to the left of $I(v)$ as well. The required
condition is $a_M(w)<\ell(v)$. Likewise, the successor and every later
member avoid $I(v)$ if and only if $r(v)<b_M(w)$. A suitable neighbor
$w$ for each nonisolated $v$ gives the PCF condition. The correspondence
between chain covers and matchings proves both directions.
\end{proof}

Some links are forced by consecutive maximum cliques.

\begin{lemma}[Forced links between maximum cliques]
\label{lem:int-forced-link}
Let $C$ and $C'$ be consecutive maximum cliques in a clique-path order.
Suppose that both have order $\omega$ and that
$C\setminus C'=\{x\}$ and $C'\setminus C=\{y\}$. If $C$ precedes
$C'$, then $x$ and $y$ have the same color in every proper
$\omega$-coloring. Moreover, $x^Ly^R$ belongs to the matching associated
with every minimum chain cover.
\end{lemma}

\begin{proof}
The common $\omega-1$ vertices have distinct colors, so both $x$ and $y$
must use the remaining color. They are nonadjacent, since otherwise
$C\cup C'$ would be a clique of order $\omega+1$. The clique-path order
places $x$ before $y$ in the interval order.

Suppose a vertex $z$ lay between them on their color chain. Every interval
in $C\cap C'$ meets both $I(x)$ and $I(y)$, so it spans the gap between
them and meets $I(z)$. The set $(C\cap C')\cup\{z\}$ would be a maximum
clique between $C$ and $C'$, contrary to their consecutiveness. Thus $x$
and $y$ are consecutive on their chain, giving the stated matching edge.
\end{proof}

\subsection{Choosing witnesses}
\label{subsec:interval-deficiency}

For some vertices, properness alone guarantees a witness. Let $\kappa(v)$
be the largest order of a clique containing $v$, and put
$\mathcal D=\{v:d(v)>0\text{ and }d(v)\geq2(\kappa(v)-1)\}$. We call
the members of $\mathcal D$ \emph{critical vertices}.

\begin{lemma}[Vertices that need no additional constraint]
\label{lem:int-automatic-vertex}
In every proper coloring, each nonisolated vertex $v$ with
$d(v)<2(\kappa(v)-1)$ has a uniquely colored neighbor.
\end{lemma}

\begin{proof}
A largest clique containing $v$ gives $\kappa(v)-1$ neighbors with
distinct colors. If none of these colors were unique in $N(v)$, each
would occur there at least twice. This would give
$d(v)\geq2(\kappa(v)-1)$, a contradiction.
\end{proof}

Choose one neighbor $f(v)$ for each $v\in\mathcal D$. We want $f(v)$ to
be the witness of $v$. From $B$, delete every edge $u^Lf(v)^R$ with
$u\in N(v)$ and every edge $f(v)^Lz^R$ with $z\in N(v)$. Denote the
remaining graph by $B_f$. These deletions prevent a neighbor of $v$
from appearing immediately before or after $f(v)$ on its color chain.

For a balanced bipartite graph $H$ with left shore $V^L$, define its Hall
deficiency by
$\operatorname{def}(H)=\max_{S\subseteq V^L}(|S|-|N_H(S)|)$.

\begin{theorem}[A matching test for chosen witnesses]
\label{thm:int-critical-deficiency}
For an interval graph $G$ of order $n$ and clique number $\omega$, the
following statements are equivalent.
\begin{enumerate}
\item $\chi_{\mathrm{pcf}}(G)=\omega$.
\item Some witness function $f$ leaves a matching of size $n-\omega$ in
$B_f$.
\item $\min_f\operatorname{def}(B_f)=\omega$.
\item Some witness function $f$ satisfies
$|N_{B_f}(S)|\geq|S|-\omega$ for every $S\subseteq V^L$.
\end{enumerate}
\end{theorem}

\begin{proof}
Suppose a PCF $\omega$-coloring is given, and link consecutive vertices of
each color class. For every critical vertex $v$, choose a witness $f(v)$.
Its predecessor and successor on its color chain cannot belong to $N(v)$.
Hence all matching edges remain in $B_f$, proving the second statement.

Conversely, a matching of size $n-\omega$ in $B_f$ gives a minimum chain
cover. The predecessor and successor of $f(v)$ are nonneighbors of $v$.
By the proof of Theorem~\ref{thm:int-chain-cover-characterization}, no
other member of this chain is adjacent to $v$. Thus $f(v)$ witnesses
$v$. Lemma~\ref{lem:int-automatic-vertex} supplies witnesses for the
remaining nonisolated vertices. Giving each chain its own color gives a
PCF $\omega$-coloring.

Hall's theorem gives $\nu(B_f)=n-\operatorname{def}(B_f)$. Since
$B_f\subseteq B$ and $\nu(B)=n-\omega$, the deficiency is at least
$\omega$. A matching of size $n-\omega$ exists if and only if this
deficiency equals $\omega$. This proves the third statement. The fourth
is the same equality written using the definition of deficiency.
\end{proof}

The edges of $B_f$ have a useful endpoint description. For each $w$, put
$\alpha_f(w)=\max(\{r(w)\}\cup\{r(v):f(v)=w\})$ and
$\beta_f(w)=\min(\{\ell(w)\}\cup\{\ell(v):f(v)=w\})$.
Then $u^Lz^R\in E(B_f)$ if and only if
$\alpha_f(u)<\ell(z)$ and $r(u)<\beta_f(z)$. The first inequality puts a
successor of $u$ to the right of every interval for which $u$ is a chosen
witness. The second puts a predecessor of $z$ to the left of every
interval for which $z$ is a chosen witness.

There are at most $\prod_{v\in\mathcal D}d(v)$ witness functions. For each
one, construct $B_f$ in $O(n^2)$ time and find a maximum matching.
Hopcroft--Karp~\cite{HopcroftKarp1973} takes $O(n^{5/2})$ time because
$B_f$ has $2n$ vertices and at most $n^2$ edges. If no witness choice
admits a matching of size $n-\omega$, the additive-one theorem gives the
optimal value $\omega+1$.

\begin{corollary}
\label{cor:int-critical-algorithm}
The PCF chromatic number can be computed in
$O(\max\{1,\Delta\}^{|\mathcal D|}n^{5/2})$ time on interval graphs.
If $\mathcal D$ is
empty, every proper $\omega(G)$-coloring is PCF.
\end{corollary}

For fixed $|\mathcal D|$, this running time is polynomial. It is FPT for
the combined parameter $|\mathcal D|+\Delta$. For $|\mathcal D|$ alone,
the bound gives an XP algorithm.

The next example shows why an arbitrary maximum matching need not give a
PCF coloring. Let $G$ have vertices $v_0,\ldots,v_5$ and right-extent
sequence $(1,2,5,3,5,5)$. Thus, for $i<j$, the edge $v_iv_j$ is present
if and only if $j\leq R_i$. Its edges are
$\{v_0v_1,v_1v_2,v_2v_3,v_2v_4,v_2v_5,v_4v_5\}$ and $\omega(G)=3$.
The chains $\{v_0,v_4\}$, $\{v_1,v_3,v_5\}$, and $\{v_2\}$ give the
PCF coloring $(0,1,2,1,0,1)$. Order both shores of $B$ by vertex index.
The corresponding matching contains the crossing edges
$v_0^Lv_4^R,v_1^Lv_3^R$. Uncrossing them gives the chains
$\{v_0,v_3,v_5\}$, $\{v_1,v_4\}$, and $\{v_2\}$. Now $v_2$ sees each
of the first two colors twice and has no witness.

Every noncrossing maximum matching fails on this graph. A matching of
size three must saturate $v_0^L,v_1^L,v_3^L$, the only left vertices with
successors. If $v_0^L$ is matched to $v_2^R$, then $v_0,v_2$ lie on the
same chain and $v_1$ has no witness. Otherwise, noncrossingness forces the
targets $v_3^R,v_4^R,v_5^R$ in this order. The vertex $v_2$ again has no
witness.

\subsection{Universal vertices and two maximal cliques}
\label{subsec:interval-joins}

Joining a universal vertex gives an exact formula for every perfect graph.
For disjoint graphs $G,H$, write $G\vee H$ for their join. The case of at
least two universal vertices below follows from the known identity
$\chi_{\mathrm{pcf}}(F\vee K_2)=\chi(F)+2$ of Ahn, Im, and Oum
\cite[Lemma~1.4]{AhnImOum2025}. We include its direct proof alongside the
one-vertex formula.

\begin{theorem}[Joining a universal clique]
\label{thm:int-universal-joins}
Let $H$ be a nonempty perfect graph, let $h=\omega(H)$, and let
$\mathcal K_h(H)$ be its family of maximum cliques. Then
\begin{equation}
\label{eq:int-one-universal}
\chi_{\mathrm{pcf}}(K_1\vee H)=
\begin{cases}
h+1,&\text{if }\displaystyle\bigcap_{Q\in\mathcal K_h(H)}Q\neq\varnothing,\\
h+2,&\text{otherwise}.
\end{cases}
\end{equation}
For every $t\geq2$, one has $\chi_{\mathrm{pcf}}(K_t\vee H)=h+t$.
\end{theorem}

\begin{proof}
Let $z$ be the new universal vertex. In a proper $(h+1)$-coloring of
$K_1\vee H$, its color is absent from $H$, while every maximum clique of
$H$ uses all other $h$ colors. The vertex $z$ witnesses every vertex of
$H$. To witness $z$, we need an $h$-coloring of $H$ with a singleton
color class, say $\{v\}$.

Such a coloring exists if and only if $\chi(H-v)\leq h-1$. Since $H$ is
perfect, this is equivalent to $\omega(H-v)\leq h-1$, which says that
$v$ belongs to every maximum clique of $H$. This proves the first case
of \eqref{eq:int-one-universal} and excludes $h+1$ colors in the second
case. In that case, color $H$ properly with $h$ colors, give $z$ a fresh
color, and recolor one vertex of $H$ with another fresh color. The first
fresh color witnesses every vertex of $H$, and the second witnesses
$z$. Thus $h+2$ colors suffice.

For $t\geq2$, the identity above applies to
$F=H\vee K_{t-2}$, with $F=H$ when $t=2$, and gives $h+t$ colors.
Directly, color $H$ properly with $h$ colors and give the vertices of
$K_t$ distinct fresh colors. Every vertex of $H$ is witnessed by a vertex
of $K_t$. Every vertex of $K_t$ is witnessed by another vertex of that
clique. The construction uses $h+t=\omega(K_t\vee H)$ colors, so it is
optimal.
\end{proof}

Universal intervals can be added to any interval model. Thus $H\vee K_2$
attains its clique bound for every interval graph $H$. Taking $H=P_3$
shows that the property $\chi_{\mathrm{pcf}}(G)=\omega(G)$ is not
hereditary within interval graphs: $P_3$ fails it, but $P_3\vee K_2$
satisfies it. Therefore no family of forbidden induced subgraphs can
characterize this property on all interval graphs.

\begin{corollary}[Two maximal cliques]
\label{cor:int-two-maximal-cliques}
Let $G$ be a connected graph with exactly two distinct maximal cliques
$A,B$. Put $a=|A|\geq b=|B|$ and $s=|A\cap B|$. Then
\begin{equation}
\label{eq:int-two-clique-formula}
\chi_{\mathrm{pcf}}(G)=
\begin{cases}
a+1,&\text{if }s=1\text{ and }a=b,\\
a,&\text{otherwise}.
\end{cases}
\end{equation}
In particular, the formula applies to interval graphs with two maximal
cliques.
\end{corollary}

\begin{proof}
Connectedness and maximality give $1\leq s<b\leq a$. An edge between
$A\setminus B$ and $B\setminus A$ would belong to a third maximal clique,
so there is no such edge. Hence
$G=K_s\vee(K_{a-s}\mathbin{\dot\cup}K_{b-s})$. For $s\geq2$,
Theorem~\ref{thm:int-universal-joins} gives the value $a$. For $s=1$, the
two exclusive cliques have an empty maximum-clique intersection if and
only if their orders agree, or equivalently $a=b$. Apply
\eqref{eq:int-one-universal}.
\end{proof}

\subsection{Clique number at most three}
\label{subsec:interval-small-clique}

For $\omega\leq3$, two maximum cliques with just one common vertex give a
complete test. The proof gives a linear-time coloring algorithm.

\begin{theorem}
\label{thm:int-small-omega}
Let $G$ be a connected interval graph containing an edge.
\begin{enumerate}
\item If $\omega(G)=2$, then $\chi_{\mathrm{pcf}}(G)=2$ if and only if
$G=K_2$; otherwise $\chi_{\mathrm{pcf}}(G)=3$.
\item If $\omega(G)=3$, then $\chi_{\mathrm{pcf}}(G)=4$ if and only if two
maximum cliques intersect in one vertex.
\end{enumerate}
An optimal coloring can be found in $O(n+m)$ time from a sorted interval
representation. Sorting arbitrary endpoints adds $O(n\log n)$ time.
\end{theorem}

\begin{proof}
Suppose first that $\omega(G)=2$. Since $G$ is chordal and triangle-free,
it is a tree. In a proper two-coloring, all neighbors of a vertex have
the opposite color. A vertex has a witness if and only if its degree is
one. Thus only $K_2$ has a PCF two-coloring.

For a PCF three-coloring of a tree, choose a root $r$ and give it color
$1$. Give one child color $2$ and all other children color $3$. At each
later vertex $z$ with parent $p$, give every child the third color,
different from $c(p)$ and $c(z)$. The root has one neighbor of color $2$.
Every other vertex has just one neighbor of its parent color. This proves
the first part.

Now let $\omega(G)=3$. Suppose maximum triangles $xab,xcd$ meet only at
$x$. In every proper three-coloring, each color different from $c(x)$
occurs at least twice in $N(x)$. The vertex $x$ has no witness, so four
colors are necessary.

Conversely, assume no pair of triangles meets in just one vertex. We first
describe the blocks of $G$. A maximal clique of order two in a chordal
graph is a bridge. Otherwise, a shortest alternative path between its ends
would give either a triangle containing the edge or an induced cycle of
length at least four. Hence every maximal clique of a 2-connected block
has order three.

Take the clique-path order $C_1,\ldots,C_s$ in such a block. A block with
$s=1$ is a triangle. If $s\geq2$, consecutive cliques meet in two
vertices, since an intersection of order one would be a cut vertex. Put
$S_i=C_i\cap C_{i+1}$. If $S_i\neq S_{i+1}$, these two edges in
$C_{i+1}$ meet in one vertex. The running-intersection property gives
$C_i\cap C_{i+2}=S_i\cap S_{i+1}$, contrary to our assumption. Thus all
$S_i$ equal a fixed edge $ab$. Every other vertex is adjacent to $a,b$,
and no two of these other vertices are adjacent. Call $a,b$ the
\emph{core} and the other vertices the \emph{tips}. Such a block is a
book; a triangle is a book with one tip. No vertex belongs to two books,
since a triangle from each would meet only in that vertex. All other
blocks are bridges.

Root the block--cut tree at a book. Give its core vertices colors $1,2$
and its tips color $3$. For a child bridge at a core, give the other
endpoint color $3$. For a child bridge at a tip, give the other endpoint
color $1$. Each core is witnessed by the other core, and each tip is
witnessed by the core of color $2$. These bridge colors do not affect the
chosen witnesses.

Continue through the block--cut tree. At an ordinary vertex $z$ reached
through a bridge $pz$, give each child the third color, different from
$c(p)$ and $c(z)$. The parent color stays unique in $N(z)$.

A nonroot book is reached through a bridge $px$. The color $X=c(x)$ has
already been assigned. Put $P=c(p)$ and let $Q$ be the third color. Use
the following rules.
\begin{itemize}
\item If $x$ is a core vertex, give the other core color $Q$ and every
tip color $P$. Give the other endpoints of child bridges at a core color
$P$, and those at a tip color $X$. The cores witness one another. Every
tip is witnessed by the core of color $Q$.
\item If $x$ is a tip, give the cores colors $P,Q$ and all tips color
$X$. Give the other endpoints of child bridges at a tip color $P$, and
those at a core color $X$. The cores witness one another. Every tip is
witnessed by the core of color $Q$. At the entry tip, its parent also has
color $P$, which leaves the witness of color $Q$ unchanged.
\end{itemize}
All bridge edges are proper. No vertex belongs to two books, so the rules
give consistent assignments. They supply a witness at every vertex and
give a PCF three-coloring.

The blocks and maximal cliques can be listed in $O(n+m)$ time. To test
for an obstructing pair, check whether every 2-connected block has a
common core edge and whether any vertex belongs to two book blocks. A
block without a common core has two different consecutive intersections
$S_i,S_{i+1}$, which give the required pair of triangles. A vertex in two
books gives a pair by choosing one triangle from each. If there is no
obstruction, the coloring rules take linear time.

When an obstruction exists, use the additive-one construction with four
colors. At most three intervals are active at any event, so direct scans
replace the heaps and balanced set by constant-time work. The
four-coloring therefore takes linear time once the endpoints are sorted.
Since a chordal graph with clique number three has $O(n)$ edges, the whole
procedure takes $O(n+m)$ time.
\end{proof}

\begin{corollary}[Small obstructions at clique number three]
\label{cor:int-omega-three-obstructions}
For an interval graph $G$ with $\omega(G)=3$, one has
$\chi_{\mathrm{pcf}}(G)=4$ if and only if $G$ contains an induced
$K_1\vee(2K_2)$ or $K_1\vee P_4$. Both graphs have five vertices, and
deleting any vertex reduces their PCF chromatic number to at most three.
\end{corollary}

\begin{proof}
Take maximum triangles $xab,xcd$ meeting only at $x$. The graph induced
by $N(x)$ is chordal and triangle-free; a triangle there would form a
$K_4$ with $x$. On $a,b,c,d$, the edges $ab,cd$ are present. There is at
most one edge between $\{a,b\}$ and $\{c,d\}$. Two such edges sharing an
endpoint would give a triangle, while two independent ones would give an
induced four-cycle. Thus these vertices induce $2K_2$ or $P_4$.
Together with $x$, they give one of the two stated graphs.

Conversely, either graph contains two maximum triangles meeting only at
their universal vertex. Theorem~\ref{thm:int-small-omega} gives the value
four. Deleting a nonuniversal vertex leaves no such pair and gives a
graph with a PCF coloring using at most three colors. Deleting the
universal vertex leaves a forest, which also needs at most three colors.
Two triangles with exactly one common vertex require at least five
vertices.
\end{proof}

\subsection{Further questions for interval graphs}
\label{subsec:interval-boundary}

The case $\omega\leq3$ is solved, but we do not know the complexity of
deciding whether an arbitrary interval graph has a PCF $\omega(G)$-coloring.
The eleven-vertex graph in
Proposition~\ref{prop:int-eleven-counterexample} shows that checking pairs
of maximum cliques is not enough. Its forced color choices prevent a
PCF four-coloring even though no pair has singleton intersection.

A related question concerns vertices common to all maximum cliques. Let
$\mathcal K_\omega(G)$ be the family of these cliques.

\begin{conjecture}[Two common vertices in maximum cliques]
\label{conj:int-maximum-core}
If a connected interval graph has at least two vertices common to all its
maximum cliques, then it has a PCF $\omega(G)$-coloring.
\end{conjecture}

The conjecture holds if every vertex belongs to a maximum clique. Choose
two common vertices $a,b$. A maximum clique containing any vertex also
contains $a,b$, so both are universal. In a proper $\omega(G)$-coloring,
their colors occur nowhere else. They witness one another, and $a$
witnesses every remaining vertex. The conjecture also holds for
$\omega\leq3$ by Theorem~\ref{thm:int-small-omega}.

Some intervals may belong only to cliques smaller than $\omega$. These
can lie at either end of the clique path or between maximum cliques. We
do not yet have a coloring argument that handles these intervals while
preserving witnesses near the common vertices.

The endpoint description of $B_f$ suggests trying to choose witnesses
and a matching together in polynomial time. It is also natural to ask
whether the $2^{O(\omega^2)}n^{O(1)}$ bound can be improved, or whether
the XP algorithm for $|\mathcal D|$ can be replaced by an FPT algorithm
for this parameter alone.

%% file: pcf_biconvex.tex
\section{Ordered bipartite graph classes}
\label{sec:ordered-bipartite-pcf}

We study three nested classes of bipartite graphs.  Bipartite permutation
graphs are biconvex, and biconvex graphs are convex.  For biconvex graphs,
we prove a tight five-color bound and give a direct exact algorithm.  A
permutation representation gives a smaller dynamic program for bipartite
permutation graphs.  For convex graphs, the PCF chromatic number is
unbounded, but we give a logarithmic upper bound and show that it is attained
for infinitely many sizes of the ordered shore.

The ordering assumptions are useful because
\textsc{PCF-$k$-Coloring} is NP-complete on general bipartite graphs for
every fixed $k\geq3$ \cite{AhnImOum2025}.  Our dynamic programs record at
most two relevant vertices of each color at a boundary.  This is enough to
distinguish the three possible neighborhood counts: zero, one, and at least
two.  Isolated vertices have no conflict-free requirement.  The empty graph
has the empty coloring and PCF chromatic number zero; the algorithms return
this answer immediately.  We assume $n\geq1$ in the running-time analyses.

\subsection{Biconvex graphs}
\label{subsec:biconvex-pcf}

A bipartite graph $G=(X,Y,E)$ is \emph{biconvex} if each shore has an order
in which every neighborhood from the other shore is an interval.  We use a
multichain ordering to describe the edges between distance layers.  For a
connected graph, let $L_0,L_1,\ldots,L_z$ be the distance layers from a
single root vertex.  The layering is a \emph{multichain ordering} if the
bipartite graph between $L_i$ and $L_{i+1}$ is a chain graph for every
$i<z$.  Thus, on either side of each boundary, the neighborhoods in the
other side are ordered by inclusion.  Every connected biconvex graph has a
multichain ordering, which can be obtained from a biconvex straight ordering
\cite{DiazDinerSernaSerra2021}.

\subsubsection{A tight five-color theorem}

The proof adapts the cyclic use of three fresh witness colors in the
AT-free argument of Sharma, Paul, and Pandey
\cite[Theorem~9, p.~507]{SharmaPaulPandey2026}. Here the multichain
ordering supplies one vertex in each layer that dominates the next layer.
This extends the known five-color bound for bipartite permutation graphs
to all biconvex graphs. We give the construction and its proof in full.

\begin{theorem}
\label{thm:ob-multichain-five}
Let $G$ be a connected bipartite graph with a multichain ordering.  Then
$\chi_{\mathrm{pcf}}(G)\leq5$.  Given the ordering, a PCF coloring with at
most five colors can be constructed in linear time.  The bound is tight,
even for bipartite permutation graphs.
\end{theorem}

\begin{proof}
Let $L_0,L_1,\ldots,L_z$ be the given ordering.  If $z=0$, then $G$ is a
single vertex and one color suffices.  Suppose $z\geq1$.  For each $i<z$,
the sets $N(u)\cap L_{i+1}$, with $u\in L_i$, are ordered by inclusion.
Their union is $L_{i+1}$, because every vertex of $L_{i+1}$ has a neighbor
in $L_i$.  A largest set in this nested family is therefore all of
$L_{i+1}$.  Choose $d_i\in L_i$ with $N(d_i)\supseteq L_{i+1}$, and
choose any $d_z\in L_z$.

Give each vertex of $L_i\setminus\{d_i\}$ color $1+(i\bmod2)$, and give
$d_i$ color $3+(i\bmod3)$.  Since $G$ is bipartite, each distance layer is
independent and every edge joins consecutive layers.  The ordinary colors
alternate between colors $1$ and $2$.  The colors of the selected vertices
cycle through $3,4,5$, and consecutive selected vertices have different
colors.  The two sets of colors are disjoint, so the coloring is proper.

The root is adjacent to $d_1$, whose color occurs only once in $L_1$.
Now consider $v\in L_i$ with $i\geq1$.  It is adjacent to $d_{i-1}$.
This is the only vertex of its color in $L_{i-1}$.  Every other possible
neighbor of $v$ lies in $L_{i+1}$.  Only $d_{i+1}$ in that layer can have
one of the colors $3,4,5$, and its color differs from that of $d_{i-1}$
because their indices differ by two modulo three.  Thus the color of
$d_{i-1}$ occurs exactly once in $N(v)$.  In the last layer there is no
$L_{i+1}$, so the same conclusion holds.  This proves the PCF condition.

To find $d_i$, choose a vertex of maximum degree into $L_{i+1}$.
Nestedness ensures that it has the required neighborhood.  One scan of the
edges computes all these degrees, and the color assignments take linear
time.  Proposition~\ref{prop:ob-bpg-tight} proves tightness.
\end{proof}

\begin{algorithm}[t]
\DontPrintSemicolon
\KwIn{A connected bipartite graph $G$ with a multichain ordering
      $L_0,L_1,\ldots,L_z$.}
\KwOut{A PCF coloring $c$ using colors from $\{1,2,3,4,5\}$.}
\If{$z=0$}{assign color $1$ to the sole vertex and \KwRet{$c$}}
\For{$i\gets0$ \KwTo $z-1$}{
  choose $d_i\in L_i$ with $N(d_i)\supseteq L_{i+1}$\;
}
choose any $d_z\in L_z$\;
\For{$i\gets0$ \KwTo $z$}{
  give every vertex of $L_i\setminus\{d_i\}$ color $1+(i\bmod2)$\;
  give $d_i$ color $3+(i\bmod3)$\;
}
\KwRet{$c$}\;
\caption{\textsc{Multichain-Five-Color}$(G,L_0,\ldots,L_z)$}
\label{alg:ob-multichain-five}
\end{algorithm}

\begin{corollary}
\label{cor:ob-biconvex-five}
Every biconvex graph has a PCF coloring with at most five colors, and the
bound is tight.
\end{corollary}

Jim\'enez et al. proved a six-color bound for biconvex graphs, with at most
three colors on each shore \cite{JimenezEtAl2026}.  The construction above
reduces the total to five without restricting the number on each shore.
Sharma, Paul, and Pandey already proved the five-color bound for bipartite
permutation graphs \cite[Corollary~5, p.~507]{SharmaPaulPandey2026}
and its tightness \cite[Lemma~1 and Fig.~2, p.~508]{SharmaPaulPandey2026}.
We recall their example and give its lower-bound argument in detail.

\begin{proposition}[Sharma et al.~{\cite[Lemma~1 and Fig.~2, p.~508]{SharmaPaulPandey2026}}]
\label{prop:ob-bpg-tight}
There is a bipartite permutation graph $H$ on twelve vertices with
$\chi_{\mathrm{pcf}}(H)=5$.
\end{proposition}

\begin{proof}
Let $U=\{u_1,u_2,u_3,u_4,u_5\}$ and
$V=\{v_1,v_2,\ldots,v_7\}$.  Set
$N(u_1)=\{v_1,v_2,v_3\}$, $N(u_2)=N(u_3)=V$, and
$N(u_4)=N(u_5)=\{v_6,v_7\}$.  In the orders
$u_1<\cdots<u_5$ and $v_1<\cdots<v_7$, the neighborhoods are intervals
whose left and right endpoints are nondecreasing.  These are strong
orders: two crossing edges complete an all-one $2\times2$ submatrix.
Hence $H$ is a bipartite permutation graph
\cite{SpinradBrandstadtStewart1987}.

Give $u_1,u_2,u_3,u_4,u_5$ the colors $1,4,1,1,1$, respectively, and give
$v_1,\ldots,v_7$ the colors $2,2,3,2,2,2,5$.  The coloring is proper.
Color $3$ witnesses $u_1$, color $5$ witnesses $u_4$ and $u_5$, and either
color witnesses $u_2$ and $u_3$.  Color $4$ appears only on $u_2$ and
witnesses every vertex of $V$.

Suppose that a PCF coloring used at most four colors.  Since
$N(v_4)=N(v_5)=\{u_2,u_3\}$, the vertices $u_2$ and $u_3$ must have
different colors.  Both are adjacent to all of $V$, so at most two colors
remain available on $V$.  The vertex $u_4$ forces $v_6$ and $v_7$ to have
different colors.  The vertex $u_1$ forces both of these available colors
to occur among $v_1,v_2,v_3$; a monochromatic neighborhood of size three
has no unique color.  Each available color therefore occurs at least twice
in $V=N(u_2)$, contradicting the PCF condition at $u_2$.  Four colors do
not suffice.
\end{proof}

\subsubsection{A direct profile dynamic program}

We now give an exact algorithm using the multichain ordering.  For each
boundary $G[L_i,L_{i+1}]$, order both shores by nonincreasing degree in
that boundary.  Write $\rho_i$ for the order on $L_i$ and
$\lambda_{i+1}$ for the order on $L_{i+1}$.  Every neighborhood is a
prefix of the order on the opposite shore.  An internal layer $L_i$ has
two orders, $\lambda_i$ toward its predecessor and $\rho_i$ toward its
successor; these orders need not agree.

Fix a positive integer $k$.  A \emph{directional profile} records the
first two occurrences of each color in an ordered layer.  For a color $a$,
its entry is $0$ if $a$ is absent, $1(x)$ if $x$ is its sole occurrence,
and $2(x,y)$ if $x,y$ are its first two occurrences in that order.  The
last entry permits further occurrences after $y$.  One order on $L_i$
has at most $(|L_i|+1)^{2k}$ profiles.  For any prefix $Q$, the profile
determines $\min\{2,|Q\cap c^{-1}(a)|\}$: it is zero if $Q$ contains
neither recorded vertex, one if it contains only the first, and two if it
contains both.  We store the endpoint of each neighborhood prefix, so this
count can be read from a profile in constant time.

A layer has no internal edges.  The following lemma therefore suffices to
check whether two profiles can describe the same coloring of a layer.

\begin{lemma}[two-profile realization]
\label{lem:ob-profile-realization}
Let $\sigma$ and $\tau$ be two orders of an independent set $W$, and let
$P,Q$ be proposed directional profiles.  Give each $w\in W$ a list
$A(w)\subseteq[k]$.  A list coloring realizing both profiles can be found,
or its nonexistence decided, in $O(k|W|)$ time.
\end{lemma}

\begin{proof}
For each color $a$, define its permitted set $U_{\sigma,a}$.  For an entry
$0$, this set is empty.  For an entry $1(x)$, it is $\{x\}$.  For an
entry $2(x,y)$, it consists of $x,y$ and every vertex after $y$ in
$\sigma$.  Define $U_{\tau,a}$ in the same way and put
$U_a=U_{\sigma,a}\cap U_{\tau,a}$.  Let $F_a$ contain all vertices
explicitly named for $a$ in either profile.

The profiles can be realized together exactly when four conditions hold.
For each color their entries must have the same type, and entries of type
$1$ must name the same vertex.  Next, $F_a\subseteq U_a$ must hold for
every $a$, and the sets $F_a$ must be pairwise disjoint.  Every vertex in
$F_a$ must allow color $a$.  Finally, each vertex outside
$\bigcup_a F_a$ must allow some color $a$ for which it belongs to $U_a$.

Any coloring realizing the profiles satisfies these conditions.  Conversely,
give every vertex in $F_a$ color $a$.  Give each remaining vertex any color
allowed by its list and permitted set.  Such a remaining vertex can only
use a color of type $2$, and it lies after the second recorded occurrence
in both orders.  It cannot change either profile.  Thus the conditions are
also sufficient.  Testing all $k$ colors at each vertex takes
$O(k|W|)$ time and constructs the coloring.
\end{proof}

We next define a test for a layer.  An $R$-profile refers to a $\rho$-order,
and a $P$-profile refers to a $\lambda$-order.  Let $\eta(H,a,Q)$ be the
capped count of color $a$ in a prefix $Q$, as determined by profile $H$.
For $v\in L_i$, profiles $R_{i-1}$ and $P_{i+1}$ give the two counts
$m_a^-(v)=\eta(R_{i-1},a,N(v)\cap L_{i-1})$ and
$m_a^+(v)=\eta(P_{i+1},a,N(v)\cap L_{i+1})$.  Set
$m_a(v)=\min\{2,m_a^-(v)+m_a^+(v)\}$.  These are the actual capped
neighbor counts for any colorings realizing the outer profiles.

If some $m_a(v)$ equals one, let $A_i(v)$ contain the colors with count
zero.  Otherwise let $A_i(v)=\varnothing$.  Thus the list is empty when
the PCF condition fails, and otherwise contains precisely the colors that
preserve properness at $v$.  Define
$C_i(R_{i-1},P_i,R_i,P_{i+1})$ to hold when $P_i,R_i$ can be realized
together using these lists.  The realization lemma tests this in
$O(k|L_i|)$ time.

After validating $L_i$, the state is the pair $(R_i,P_{i+1})$.
A transition from $(R_{i-1},P_i)$ to $(R_i,P_{i+1})$ is allowed when
$C_i$ holds.  At the first layer, use its sole order for both internal
profiles and initialize the states satisfying
$C_0(\varnothing,R_0,R_0,P_1)$.  At the last layer, accept a state when
$C_z(R_{z-1},P_z,P_z,\varnothing)$ holds.  A missing neighboring layer
contributes zero to every count.  Repeating the sole order at an endpoint
only checks the coloring of that endpoint layer.

\begin{algorithm}[t]
\DontPrintSemicolon
\KwIn{A connected bipartite graph $G$, a multichain ordering
      $L_0,\ldots,L_z$, and a positive integer $k$.}
\KwOut{A PCF $k$-coloring, or \textsc{No}.}
\If{$z=0$}{assign any color to the sole vertex and \KwRet{the coloring}}
construct all orders $\rho_i,\lambda_i$ and neighborhood-prefix endpoints\;
enumerate the directional profiles of each ordered layer\;
$\mathcal T\gets\varnothing$\;
\ForEach{pair $(R_0,P_1)$ satisfying
  $C_0(\varnothing,R_0,R_0,P_1)$}{
  insert $(R_0,P_1)$ into $\mathcal T$\;
}
\For{$i\gets1$ \KwTo $z-1$}{
  $\mathcal T'\gets\varnothing$\;
  \ForEach{$(R_{i-1},P_i)\in\mathcal T$ and every pair
    $(R_i,P_{i+1})$}{
    \If{$C_i(R_{i-1},P_i,R_i,P_{i+1})$ holds}{
      insert $(R_i,P_{i+1})$ into $\mathcal T'$ and store a predecessor\;
    }
  }
  $\mathcal T\gets\mathcal T'$\;
}
\ForEach{$(R_{z-1},P_z)\in\mathcal T$}{
  \If{$C_z(R_{z-1},P_z,P_z,\varnothing)$ holds}{
    trace the stored predecessors to recover all layer profiles\;
    realize each layer using Lemma~\ref{lem:ob-profile-realization}\;
    \KwRet{the combined coloring}\;
  }
}
\KwRet{\textnormal{\textsc{No}}}\;
\caption{\textnormal{\textsc{Fixed-PCF-Multichain}}$(G,k)$}
\label{alg:ob-fixed-multichain}
\end{algorithm}

\begin{theorem}
\label{thm:ob-fixed-multichain}
For each fixed positive integer $k$,
\textnormal{\textsc{PCF-$k$-Coloring}} can be solved on an $n$-vertex
bipartite graph supplied with a multichain ordering in $O_k(n^{8k+1})$
time. An accepting coloring can be recovered within the same time bound
by retaining predecessor pointers.
\end{theorem}

\begin{proof}
A PCF $k$-coloring gives a profile in each order.  All neighbors of a vertex
of $L_i$ belong to $L_{i-1}\cup L_{i+1}$.  The outer profiles therefore
give its true capped neighbor counts.  Its own color has count zero, and
some color has count one.  Its color belongs to $A_i(v)$, so every layer
passes the test.  The coloring gives an accepting sequence of states.

Conversely, take an accepting sequence.  Realize the two profiles of each
layer using the realization lemma and its lists.  Each layer coloring has
exactly the profiles used in the tests on its adjacent layers.  Consequently,
the neighbor counts used to define every list agree with the capped counts
in the combined coloring.  A vertex receives a color with neighbor count
zero and has some neighbor color with count one.  The combined coloring is
therefore proper and conflict-free.

At layer $i$, a transition chooses two profiles of $L_i$ and one profile
of each adjacent layer.  There are at most
$(|L_{i-1}|+1)^{2k}(|L_i|+1)^{4k}(|L_{i+1}|+1)^{2k}$ choices.
Each test takes $O(k|L_i|)$ time, so the total is at most
$O(kn(n+1)^{8k})=O_k(n^{8k+1})$.  This bound includes the smaller endpoint
tests and the polynomial preprocessing.

For reconstruction, retain a predecessor for each reachable state at every
layer. Trace an accepting path and apply the realization lemma once per
layer. This takes $O(kn)$ additional time and returns the coloring within
the stated time bound.
\end{proof}

\begin{lemma}
\label{lem:ob-bipartite-one-two}
A nonempty bipartite graph $G$ has $\chi_{\mathrm{pcf}}(G)=1$ exactly when
it is edgeless.  If it has an edge, then $\chi_{\mathrm{pcf}}(G)=2$ exactly
when $\Delta(G)\leq1$.
\end{lemma}

\begin{proof}
One color is proper exactly when there is no edge.  Suppose $G$ has an
edge.  In a proper two-coloring of a nontrivial connected bipartite
component, all neighbors of a vertex have the opposite color.  That color
occurs once exactly when the vertex has degree one.  Thus a proper
two-coloring is PCF exactly when every component containing an edge is a
single edge.
\end{proof}

\begin{algorithm}[htbp!]
\DontPrintSemicolon
\KwIn{A biconvex graph $G$.}
\KwOut{$\chi_{\mathrm{pcf}}(G)$ and an optimal coloring.}
\If{$V(G)=\varnothing$}{\KwRet{$0$ and the empty coloring}}
\If{$E(G)=\varnothing$}{\KwRet{$1$ and the one-coloring}}
\If{$\Delta(G)\leq1$}{\KwRet{$2$ and a proper two-coloring}}
compute a multichain ordering of every nontrivial component\;
run Algorithm~\ref{alg:ob-fixed-multichain} with $k=3$ on every component\;
\If{all components accept}{combine their colorings and \KwRet{$3$}}
run Algorithm~\ref{alg:ob-fixed-multichain} with $k=4$ on every component\;
\If{all components accept}{combine their colorings and \KwRet{$4$}}
apply Algorithm~\ref{alg:ob-multichain-five} componentwise and reuse the palette\;
\KwRet{$5$ and the resulting coloring}\;
\caption{\textsc{Exact-PCF-Biconvex}$(G)$}
\label{alg:ob-exact-biconvex}
\end{algorithm}

\begin{theorem}
\label{thm:ob-exact-biconvex}
The PCF chromatic number of an $n$-vertex biconvex graph and an optimal
coloring can be computed in $O(n^{33})$ time.
\end{theorem}

\begin{proof}
The lemma settles the values one and two.  If neither case applies, the
value is at least three.  The calls with $k=3$ and $k=4$ decide whether
three or four colors suffice.  If both reject, at least five colors are
needed, and Corollary~\ref{cor:ob-biconvex-five} supplies five.
The call with $k=4$ takes $O(n^{8\cdot4+1})=O(n^{33})$ time and dominates
the remaining work.  Components can use the same palette, so their optimum
is the maximum of their individual optima.  The sum of their running times
is bounded by $O(n^{33})$ as well.
\end{proof}

\subsection{Bipartite permutation graphs}
\label{subsec:bipartite-permutation-pcf}

A permutation representation gives smaller states.  For each color we keep
two extreme vertices before the current cut and two after it.  The
fixed-palette algorithm below applies to every permutation graph.  Together
with the five-color bound, it gives an $O(n^{17})$ exact algorithm for
bipartite permutation graphs.

A permutation diagram represents vertices by segments with endpoints on two
parallel lines; no two endpoints on the same line coincide.  Order the
vertices $v_1,\ldots,v_n$ by their upper endpoints, and let $p(v)$ be the
rank of the lower endpoint of $v$.  For $i<j$, vertices $v_i,v_j$ are
adjacent exactly when $p(v_i)>p(v_j)$.

A bipartite graph is a bipartite permutation graph exactly when its shores
have \emph{strong orders} $<_X,<_Y$: if $x<_Xx'$ and $y<_Yy'$ and the
edges $xy',x'y$ exist, then $xy,x'y'$ exist too.  Such graphs can be
recognized and their strong orders found in linear time
\cite{SpinradBrandstadtStewart1987}.  A corresponding permutation diagram
can be constructed in polynomial time, within the running-time bound of
our dynamic program.

At the cut after $v_t$, put $A_t=\{v_1,\ldots,v_t\}$ and
$B_t=V(G)\setminus A_t$.  For a set $S$, let
$\operatorname{high}_2(S)$ contain its two vertices with largest lower
positions, or all of $S$ when $|S|\leq2$.  Define
$\operatorname{low}_2(S)$ using the two smallest lower positions.

\begin{lemma}[two-extremes lemma]
\label{lem:ob-two-extremes}
For every cut $(A_t,B_t)$, every $S\subseteq A_t$, and every $b\in B_t$,
we have $\min\{2,|N(b)\cap S|\}=|N(b)\cap\operatorname{high}_2(S)|$.
For every $S\subseteq B_t$ and every $a\in A_t$, we have
$\min\{2,|N(a)\cap S|\}=|N(a)\cap\operatorname{low}_2(S)|$.
\end{lemma}

\begin{proof}
For $a\in A_t$ and $b\in B_t$, adjacency means $p(a)>p(b)$.
The neighbors of a fixed $b$ in $A_t$ form a suffix in the lower-endpoint
order.  If this suffix contains at least two members of $S$, it contains
the two highest members.  If it contains one member, that member is also
retained.  This proves the first identity.  The neighbors of a fixed $a$
in $B_t$ form a prefix, so retaining the two lowest members proves the
second identity.
\end{proof}

\subsubsection{The fixed-palette dynamic program}

Fix a positive integer $k$.  A state at cut $t$ is
$(\mathcal R;\mathcal L)$, where
$\mathcal R=(R_1,\ldots,R_k)$ and $\mathcal L=(L_1,\ldots,L_k)$.
Each $R_i\subseteq A_t$ and $L_i\subseteq B_t$ has size at most two.
The sets within each tuple must be pairwise disjoint.  The tuple
$\mathcal R$ summarizes the colors already assigned, and $\mathcal L$
is a proposed summary of the colors still to be assigned.  We call the
latter tuple a promise.

The state $D_t(\mathcal R;\mathcal L)$ is true if there is a coloring
$f:A_t\to[k]$ with two properties.  First,
$R_i=\operatorname{high}_2(f^{-1}(i))$ for each $i$.  Second, every
processed vertex is valid with respect to the promise.  To state the latter
condition, for $x\in A_t$ set
$q_i(x)=\min\{2,|\{y\in N(x)\cap A_t:f(y)=i\}|+|N(x)\cap L_i|\}$.
We require $q_{f(x)}(x)=0$ and, unless $x$ is isolated, $q_i(x)=1$ for
some $i$.  These conditions express properness and the existence of a
unique neighbor color.  A promise need not itself have a full realization
at this cut.  An accepting sequence of transitions will ensure its
consistency with the colors eventually assigned.

At $t=0$, mark
$D_0((\varnothing,\ldots,\varnothing);\mathcal L)$ true for every
admissible promise.  At a later step, put $v=v_t$.  Choose a previous
processed tuple $\mathcal R^-$, a new promise $\mathcal L$, and a color
$\gamma$ for $v$.  Define the previous promise by
$L^-_\gamma=\operatorname{low}_2(L_\gamma\cup\{v\})$ and
$L^-_i=L_i$ for $i\neq\gamma$.
Only continue if $D_{t-1}(\mathcal R^-;\mathcal L^-)$ is true.
For each $i$, compute
$d_i=\min\{2,|N(v)\cap R_i^-|+|N(v)\cap L_i|\}$.
The transition is allowed if $d_\gamma=0$ and, unless $v$ is isolated,
some $d_i=1$.  Set
$R_\gamma=\operatorname{high}_2(R^-_\gamma\cup\{v\})$ and
$R_i=R_i^-$ for the other colors, and mark
$D_t(\mathcal R;\mathcal L)$ true.

\begin{algorithm}[htbp!]
\DontPrintSemicolon
\KwIn{A permutation graph $G$ with diagram $v_1,\ldots,v_n$, and a positive
      integer $k$.}
\KwOut{A PCF $k$-coloring, or \textsc{No}.}
\If{$V(G)=\varnothing$}{\KwRet{the empty coloring}}
mark $D_0((\varnothing,\ldots,\varnothing);\mathcal L)$ true for every
admissible promise tuple $\mathcal L$\;
\For{$t\gets1$ \KwTo $n$}{
  $v\gets v_t$ and clear table $D_t$\;
  \ForEach{processed tuple $\mathcal R^-$, new promise $\mathcal L$, and
           color $\gamma\in[k]$}{
    form $\mathcal L^-$ by adding $v$ to color $\gamma$ and retaining
    the two lowest positions\;
    \If{$D_{t-1}(\mathcal R^-;\mathcal L^-)$ is true}{
      compute $d_i=\min\{2,|N(v)\cap R_i^-|+|N(v)\cap L_i|\}$ for all $i$\;
      \If{$d_\gamma=0$ and either $v$ is isolated or some $d_i=1$}{
        add $v$ to $R^-_\gamma$ and retain its two highest positions to
        form $\mathcal R$\;
        mark $D_t(\mathcal R;\mathcal L)$ true and store a predecessor\;
      }
    }
  }
}
\If{$D_n(\mathcal R;(\varnothing,\ldots,\varnothing))$ is true for some
      $\mathcal R$}{
  trace the predecessors and \KwRet{the recovered coloring}\;
}
\KwRet{\textnormal{\textsc{No}}}\;
\caption{\textnormal{\textsc{Fixed-PCF-Permutation}}$(G,k)$}
\label{alg:ob-fixed-permutation}
\end{algorithm}

\begin{lemma}
\label{lem:ob-permutation-dp-correct}
Algorithm~\ref{alg:ob-fixed-permutation} accepts exactly the permutation
graphs that have a PCF $k$-coloring.
\end{lemma}

\begin{proof}
We prove the stated meaning of $D_t$ by induction.  At $t=0$ there are no
processed vertices, so every promise is valid.  Consider an allowed
transition assigning $\gamma$ to $v_t$.  Updating $\mathcal R^-$ retains
the correct two highest vertices of each processed color class.

For an old vertex $x\in A_{t-1}$, only the contribution of color $\gamma$
changes.  In the previous state its promised contribution comes from
$\operatorname{low}_2(L_\gamma\cup\{v_t\})$.  By the second identity in
Lemma~\ref{lem:ob-two-extremes}, this has the same capped contribution as
$L_\gamma\cup\{v_t\}$.  Thus moving $v_t$ from the promise to the
processed side preserves the full capped neighbor-count vector of $x$.
The old vertices remain valid.

For $v_t$, the first identity in the lemma shows that $R_i^-$ gives the
correct capped contribution from the processed color-$i$ vertices.  Adding
its promised contribution from $L_i$ gives $d_i$.  The transition therefore
checks exactly the two required conditions at $v_t$.  Every state produced
by a transition has the stated meaning.

For the converse, take a coloring $f$ satisfying the meaning of a state at
$t$.  Remove $v_t$, put $\gamma=f(v_t)$, and let $\mathcal R^-$ be the
correct summaries of the remaining color classes.  Form $\mathcal L^-$
by the prescribed update from $\mathcal L$.  The same capped-count
identities show that the restricted coloring satisfies the meaning of this
previous state.  It is present by induction, and the counts at $v_t$
allow the transition back to the chosen state.  Hence every state with the
stated meaning is generated.  At $t=n$, all promise sets are empty, so the
local conditions are precisely those of a PCF coloring of the whole graph.
\end{proof}

\begin{theorem}
\label{thm:ob-permutation-fixed-k}
Given a permutation diagram of an $n$-vertex graph,
\textnormal{\textsc{PCF-$k$-Coloring}} can be solved in
$O(k^2n^{4k+1})$ time for each fixed positive $k$. Retaining predecessor
pointers recovers a coloring within the same time bound.
\end{theorem}

\begin{proof}
For one color, a summary has $1+n+\binom n2$ possibilities.  For $n\geq2$
this is at most $n^2$.  Thus the $k$ processed and $k$ promised summaries
give at most $n^{4k}$ states at a cut.  At each step we enumerate at most
$n^{2k}$ previous processed tuples, $n^{2k}$ new promises, and $k$ choices
of $\gamma$.  Updating the tuples and checking all counts takes $O(k)$
time.  Adjacency is tested in constant time from the endpoint ranks.
Over $n$ steps, the total is $O(k^2n^{4k+1})$; the one-vertex case can
be answered directly.

To recover a coloring within the same running time, keep one predecessor
and the assigned color for each reachable state at every cut. Trace back
from any accepting state. No repeated runs of the decision algorithm are
needed for this recovery.
\end{proof}

\subsubsection{Computing the exact value}

Sharma et al. established the five-color bound for bipartite
permutation graphs \cite[Corollary~5, p.~507]{SharmaPaulPandey2026};
its tightness is proved in their Lemma~1 on p.~508. It also follows
from Corollary~\ref{cor:ob-biconvex-five} and the recalled example.
We combine this range with the fixed-palette algorithm.

\begin{algorithm}[t]
\DontPrintSemicolon
\KwIn{A bipartite permutation graph $G$.}
\KwOut{$\chi_{\mathrm{pcf}}(G)$ and an optimal coloring.}
\If{$V(G)=\varnothing$}{\KwRet{$0$ and the empty coloring}}
\If{$E(G)=\varnothing$}{\KwRet{$1$ and the one-coloring}}
\If{$\Delta(G)\leq1$}{\KwRet{$2$ and a proper two-coloring}}
construct a permutation diagram of $G$\;
run Algorithm~\ref{alg:ob-fixed-permutation} with $k=3$\;
\If{it accepts}{\KwRet{$3$ and the recovered coloring}}
run Algorithm~\ref{alg:ob-fixed-permutation} with $k=4$\;
\If{it accepts}{\KwRet{$4$ and the recovered coloring}}
construct five-colorings componentwise using
Algorithm~\ref{alg:ob-multichain-five} and reuse the palette\;
\KwRet{$5$ and that coloring}\;
\caption{\textsc{Exact-PCF-Bipartite-Permutation}$(G)$}
\label{alg:ob-exact-bpg}
\end{algorithm}

\begin{theorem}
\label{thm:ob-exact-bpg}
The PCF chromatic number of an $n$-vertex bipartite permutation graph and
an optimal coloring can be computed in $O(n^{17})$ time.
\end{theorem}

\begin{proof}
Lemma~\ref{lem:ob-bipartite-one-two} decides the values one and two.
The fixed-palette calls for $k=3$ and $k=4$ take $O(n^{13})$ and
$O(n^{17})$ time, respectively.  If both reject, five colors are necessary,
and Corollary~\ref{cor:ob-biconvex-five} constructs a coloring with five.
The call with $k=4$ dominates the running time.
\end{proof}

\subsection{Convex bipartite graphs}
\label{subsec:convex-bipartite-pcf}

Let $G=(X,Y,E)$ be convex over $X=(x_1,\ldots,x_p)$, so each
$N(y)$ is an interval in this order.  The PCF chromatic number is
unbounded on convex bipartite graphs \cite{JimenezEtAl2026}.  We give a
logarithmic bound in $p$ and show that its additive constant is best possible.

\begin{theorem}
\label{thm:ob-convex-bound}
If $G=(X,Y,E)$ is convex over an ordered shore $X$ of size $p$, then
$\chi_{\mathrm{pcf}}(G)\leq\lceil\log_2(p+1)\rceil+3$.  Given the
convex order, a coloring meeting this bound can be constructed in
$O(|V(G)|+|E(G)|)$ time.
\end{theorem}

\begin{proof}
Put $h=\lceil\log_2(p+1)\rceil$.  For $1\leq i\leq p$, give $x_i$
color $1+\nu_2(i)$, where $\nu_2(i)$ is the exponent of the largest power
of two dividing $i$.  Since $p<2^h$, this uses at most $h$ colors.
Every nonempty interval of indices has a unique maximum color.  Indeed,
between any two indices with the same $2$-adic valuation there is an index
with a larger valuation.  Two indices therefore cannot both realize the
maximum valuation in an interval.  Each nonisolated $y\in Y$ has an
interval neighborhood, so this maximum color witnesses $y$.

We must also provide a witness for each nonisolated vertex of $X$.  We use
the standard three-color construction for interval families, recalled by
Aronov et al. \cite{AronovDeBergMarkovicWoeginger2020}.
Regard the nonempty sets $N(y)$ as intervals of $X$.  Choose a
deletion-minimal subfamily $\mathcal S$ with the same union as the whole
family.  No three selected intervals contain a common point.  To see this,
among three such intervals choose one with the leftmost left endpoint and
one with the rightmost right endpoint.  If they are distinct, they
intersect and together cover the third.  If they are the same interval,
it contains the other two.  Either case makes a selected interval
redundant, contrary to minimality.

The intersection graph of $\mathcal S$ is therefore a triangle-free
interval graph.  Interval graphs are chordal, so this graph is a forest.
Properly color this forest with colors $\alpha,\beta$, and give every
vertex of $Y$ whose interval is outside $\mathcal S$ a third color
$\gamma$.  Each nonisolated point of $X$ belongs to one or two selected
intervals.  If it belongs to one, that interval's color appears once among
its neighbors.  If it belongs to two, those intervals intersect and have
different colors, each appearing once.  The vertices colored $\gamma$
do not affect either unique selected color.

Use $h+1,h+2,h+3$ for $\alpha,\beta,\gamma$.  The shores now use
disjoint palettes, so the coloring is proper.  The coloring of $X$ witnesses
all nonisolated vertices of $Y$, and the selected intervals witness all
nonisolated vertices of $X$.  Hence the coloring is PCF.  This also covers
$p=0$, when $G$ is edgeless.

For the running time, store the two endpoints of each nonempty neighborhood
and compute the number of intervals covering each point.  Initially retain
all intervals and scan them once.  Delete an interval if every point in it
has coverage at least two, decreasing the coverage counters on its points.
An interval indispensable when scanned remains indispensable after later
deletions, so the final family is deletion-minimal.  Each incidence is
examined only constantly many times, for $O(|E(G)|)$ work.

Bucket the selected intervals by their integer left endpoints and sweep
from left to right.  At most one earlier selected interval can still be
active when the next starts, since the selected family has depth at most
two.  The sweep constructs the intersection forest and its two-coloring in
linear time.  Finally, the colors $1+\nu_2(i)$ can be generated in $O(p)$
total time: odd positions receive color $1$, and an even position $i$
receives one plus the color at position $i/2$.  All remaining assignments
are linear.
\end{proof}

\begin{algorithm}[htbp!]
\DontPrintSemicolon
\KwIn{A convex bipartite graph $G=(X,Y,E)$ and an order
      $X=(x_1,\ldots,x_p)$ witnessing convexity.}
\KwOut{A PCF coloring with at most $\lceil\log_2(p+1)\rceil+3$ colors.}
$h\gets\lceil\log_2(p+1)\rceil$\;
\For{$i\gets1$ \KwTo $p$}{give $x_i$ color $1+\nu_2(i)$\;}
$\mathcal S\gets\{y\in Y:N(y)\neq\varnothing\}$\;
compute the coverage count $\operatorname{cov}(x)$ of each $x\in X$\;
\ForEach{$y\in\mathcal S$ in a fixed order}{
  \If{$\operatorname{cov}(x)\geq2$ for every $x\in N(y)$}{
    remove $y$ from $\mathcal S$ and decrease every counter on $N(y)$\;
  }
}
construct the intersection forest of the intervals of $\mathcal S$\;
properly color this forest with $h+1,h+2$\;
give each vertex of $Y\setminus\mathcal S$ color $h+3$\;
\KwRet{the combined coloring}\;
\caption{\textsc{Convex-PCF-Color}$(G,X)$}
\label{alg:ob-convex-color}
\end{algorithm}

The additive constant three is attained for infinitely many values of $p$.

\begin{theorem}
\label{thm:ob-convex-sharp}
For each integer $r\geq3$, there is a convex bipartite graph $G_r$ with
$|X|=2^{r-1}+3$ and $\chi_{\mathrm{pcf}}(G_r)=r+3$.  Thus equality holds
in Theorem~\ref{thm:ob-convex-bound} for infinitely many ordered-shore sizes.
\end{theorem}

\begin{proof}
Let $C=(c_1,\ldots,c_{2^{r-1}})$ and order the shore $X$ as
$(\ell,c_1,\ldots,c_{2^{r-1}},m,z)$.  For each nonempty interval $J$ of
$C$, introduce a vertex $b_J\in Y$ with $N(b_J)=J$.  Add vertices
$a_1,a_2,d_1,d_2$ to $Y$.  Set
$N(a_1)=N(a_2)=\{\ell\}\cup C\cup\{m\}$ and
$N(d_1)=N(d_2)=C\cup\{m,z\}$.  All these neighborhoods are intervals in
$X$, so $G_r$ is convex.

We first use an elementary bound for color words.  If every nonempty
contiguous subword has a uniquely occurring color and the whole word uses
at most $t$ colors, its length is at most $2^t-1$.  For induction, choose a color
occurring once in the whole word.  The subwords on its two sides have the
same property and use at most $t-1$ colors.  Each has length at most
$2^{t-1}-1$, giving the required bound after adding the middle position.
The empty word supplies the base case $t=0$.

The neighborhood constraint at every $b_J$ gives this property to the color
word on $C$.  Since $|C|=2^{r-1}$, the word needs at least $r$ colors:
$r-1$ colors could cover at most $2^{r-1}-1$ positions.

Suppose the whole graph had a PCF coloring with at most $r+2$ colors.
Each of $a_1,a_2,d_1,d_2$ is adjacent to all of $C$, so none can use a
color appearing on $C$.  At most two colors remain for these four vertices.
The vertex $\ell$ has neighborhood $\{a_1,a_2\}$ and forces their colors
to differ.  Similarly, $z$ forces different colors on $d_1,d_2$.
Consequently, the same two available colors each occur twice in
$N(m)=\{a_1,a_2,d_1,d_2\}$.  The PCF condition fails at $m$, a
contradiction.  At least $r+3$ colors are needed.

For $r\geq3$, the equality $|X|=2^{r-1}+3$ gives
$\lceil\log_2(|X|+1)\rceil=r$.  The upper bound from
Theorem~\ref{thm:ob-convex-bound} therefore supplies exactly $r+3$ colors.
\end{proof}

In terms of the total number $n$ of vertices, the upper bound implies
$\chi_{\mathrm{pcf}}(G)\leq\lceil\log_2(n+1)\rceil+3$.
The nested dyadic graphs $H_k$ of Jim\'enez et al. have $2^{k+1}-1$
vertices and PCF chromatic number at least
$k+2=\log_2(|V(H_k)|+1)+1$ \cite[Proposition~3.8]{JimenezEtAl2026}.
Their neighborhoods on the ordered shore are dyadic intervals, so these
graphs are convex, in addition to the bipartite circle representation stated
there.  Adding isolated vertices fills the gaps between these sizes.
It follows that the largest PCF chromatic number among convex bipartite
graphs on $n$ vertices is $\log_2 n+O(1)$.
Algorithm~\ref{alg:ob-convex-color} constructs a coloring meeting the upper
bound.  The complexity of computing the exact PCF chromatic number on
convex bipartite graphs remains open.